%% file: main.tex
\documentclass[acmtog]{acmart}
\acmSubmissionID{372}

\usepackage{booktabs} 

\usepackage{paralist}
\usepackage{textcomp}
\usepackage{microtype}

\usepackage{times}
\usepackage{epsfig}
\usepackage{graphicx}
\usepackage{float}
\usepackage{wrapfig}
\usepackage{amsmath,amsthm}
\usepackage{subfig}
\usepackage{multirow}

\usepackage{pgfplots}
\pgfplotsset{compat=1.17}

\usepackage{setspace}

\setcopyright{acmcopyright}\acmJournal{TOG}
\acmYear{2022}\acmVolume{41}\acmNumber{6}\acmArticle{258}\acmMonth{12} \acmDOI{10.1145/3550454.3555492}

\newcommand\mypara[1]{\vspace{1mm}\noindent\textbf{#1}}
\usepackage{mathtools}
\DeclarePairedDelimiter\floor{\lfloor}{\rfloor}

\newcommand{\PD}[2]{\frac{\partial{#1}}{\partial{#2}}}

\usepackage[ruled]{algorithm2e} 

\SetAlFnt{\small}
\SetAlCapFnt{\small}
\SetAlCapNameFnt{\small}
\SetAlCapHSkip{0pt}

\acmJournal{TOG}

\begin{document}
\title{Differentiable Hybrid Traffic Simulation}

\author{Sanghyun Son}
\affiliation{%
 \institution{University of Maryland, College Park}
 \country{USA}}
\email{shh1295@umd.edu}

\author{Yi-Ling Qiao}
\affiliation{%
 \institution{University of Maryland, College Park}
 \country{USA}}
\email{yilingq@umd.edu}

\author{Jason Sewall}
\affiliation{%
 \institution{NVIDIA}
 \country{USA}
}
\email{jasonsewall@gmail.com}

\author{Ming C. Lin}
\affiliation{%
\institution{University of Maryland, College Park}
\country{USA}}
\email{lin@umd.edu}

\begin{spacing}{0.994}
\begin{abstract}
We introduce a novel {\em differentiable hybrid traffic simulator}, which simulates traffic using a hybrid model of both macroscopic and microscopic models and can be directly {\em integrated into a neural network for traffic control and flow optimization}. This is the first differentiable traffic simulator for macroscopic and hybrid models that can compute gradients for traffic states across time steps and inhomogeneous lanes. To compute the gradient flow between two types of traffic models in a hybrid framework, we present a novel intermediate conversion component that bridges the lanes in a differentiable manner as well. We also show that we can use analytical gradients to accelerate the overall process and enhance scalability. Thanks to these gradients, our simulator can provide more efficient and scalable solutions for complex learning and control problems posed in traffic engineering than other existing algorithms. Refer to \textcolor{magenta}{https://sites.google.com/umd.edu/diff-hybrid-traffic-sim} for our project.
\end{abstract}


%
%


\maketitle

\section{Introduction}
\label{sec:introduction}
\input{tex/1_introduction}

\section{Previous Work}
\label{sec:previous}
\input{tex/2_previous}

\section{Macroscopic Model}
\label{sec:macro}
\input{tex/3_macro}

\section{Microscopic Model}
\label{sec:micro}
\input{tex/4_micro}

\section{Macro-Micro-Macro Conversion}
\label{sec:conversion}
\input{tex/5_conversion}

\section{Experimental Results}
\label{sec:experiments}
\input{tex/6_experiments}

\section{Conclusions}
\label{sec:conclusions}

\input{tex/conclusions}


\bibliographystyle{ACM-Reference-Format}
\bibliography{main}

\newpage
\appendix
\onecolumn
\section{Solution to the Riemann problem in ARZ model}
\label{appendix-riemann-solution}
\input{tex/appendix_riemann_solution}

\section{Differentiating IDM}
\label{appendix-idm}
\input{tex/appendix_idm}

\end{spacing}

\end{document}

%% file: tex/1_introduction.tex
Automobile traffic is a dynamic phenomenon that emerges from the interactions of a multitude of distinct entities, and its complexity has led analysts to rely on simulations to solve complex real-world traffic problems, such as signal control, congestion, road design, and urban planning~\cite{lieberman1997traffic}.  Traffic simulation has become increasingly more important due to population growth, fast-growing technologies, and demands for autonomous driving. Autonomous driving agents need accurate and efficient models to simulate all types of plausible traffic scenarios to capture corner cases and accelerate learning-based training that requires data not easily available from real-world capturing~\cite{akhauri2020enhanced}. 

Techniques for traffic simulation can be broadly classified as either \emph{macroscopic} or \emph{microscopic} based on their modeling assumptions. Macroscopic models describe traffic evolution as a system of partial differential equations (PDE), and the traffic state is represented as a {\em continuum} of values across the road network. Since we treat vehicles as particles translated by convection under this assumption, it is computationally efficient but provides coarse simulation results. In contrast, microscopic models represent traffic through individual vehicles, or agents, that are evolved individually and which together characterize the traffic state of the road network~\cite{sewall2011interactive}. Therefore, this approach gives us fine-grained details but often requires much more computation resources than macroscopic models.

\begin{figure}
\centering
\includegraphics[scale=0.17125]{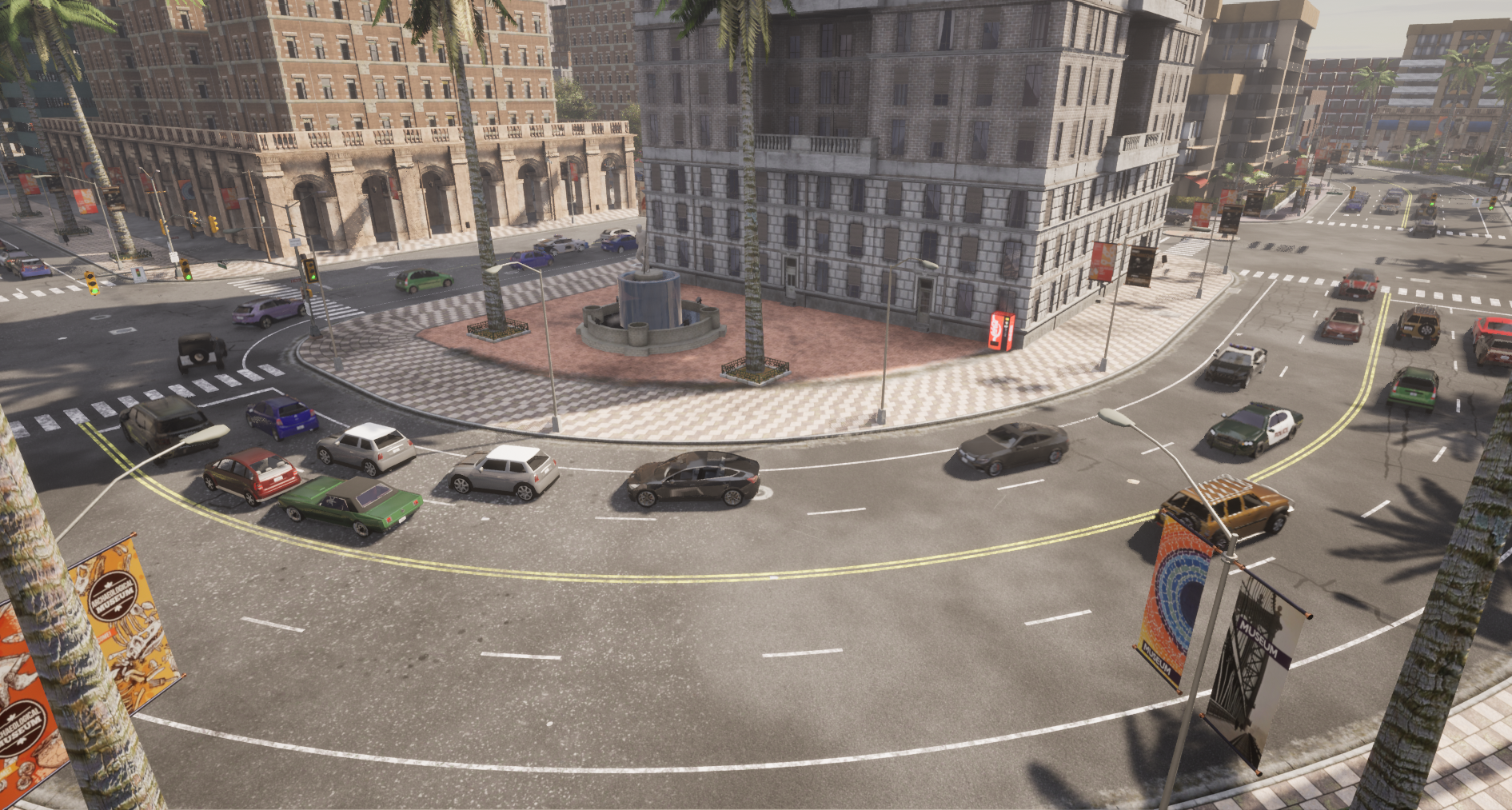}
\vspace{-2em}
\caption{ {\bf Traffic simulation in an urban environment.} Traffic simulation can be used to analyze complex traffic dynamics. Among two popular traffic models, by simulating areas of interest with the microscopic one and using macroscopic simulation elsewhere, we can reduce the overall computational cost without compromising significant details. For instance, the center of intersection can be simulated with discrete, agent-based models for higher-fidelity inter-vehicle dynamic interaction. ({\em This scene is rendered in CARLA}~\cite{Dosovitskiy17}.)}
\vspace{-1.5em}
\label{fig:intro-traffic-sim}
\end{figure}

In this work, we propose to adopt both microscopic and macroscopic models to create a more general traffic simulator that is differentiable (Figure~\ref{fig:intro-traffic-sim}). This kind of hybrid approach has already been proven to simulate large-scale traffic environments much more efficiently than either of the approaches while maintaining integrity~\cite{sewall2011interactive}. Here we leverage the power of the hybrid approach to support large-scale traffic scenes, which will be demonstrated in our experiments. During simulation, our framework computes gradient information that provides abundant insight into traffic dynamics. This approach results in enhanced sample efficiency in our simulation, as each sample of traffic simulation comes with gradient information explaining how and why such an event has occurred. Moreover, this differentiability allows us to integrate our framework with neural networks to support a wide range of traffic control, planning, management, and
flow optimization problems. 

We first derive the analytical formulation of the differentiable traffic models for both macroscopic and microscopic paradigms. The resulting representations can simulate 
traffic on road networks for any metropolitan region more efficiently. Furthermore, we present 
a probabilistic method to convert traffic states between macroscopic and microscopic ones, which is also differentiable. With this novel differentiable traffic simulation model, we can find near-optimal solutions to some traffic problems that were otherwise not previously solvable, and we can do so efficiently. To summarize, 
in this paper we introduce the following main results: 
\begin{compactitem}
\item Derivation of a differentiable macroscopic traffic model based on ARZ model~\cite{aw2000resurrection, zhang2002non}
 (Sec.~\ref{sec:macro});
\item Analytical formulation of differentiable Intelligent Driver Model 

~\cite{treiber2000congested} for agent-based vehicle dynamics (Sec.~\ref{sec:micro});
\item Differentiable conversion between deterministic and probabilistic instantiation of two traffic models (Sec.~\ref{sec:conversion});
\item Application of the first differentiable hybrid traffic simulation framework to traffic control (Sec.~\ref{sec:experiments}).
\end{compactitem}
In our experiments, we observe up to {\em an order of magnitude} speedup in runtime performance, in both forward simulation and back propagation process, compared to a baseline differentiable simulator based on automatic differentiation.

%% file: tex/2_previous.tex

\subsection{Traffic models}
The preponderance of macroscopic models are based on the \emph{density} and \emph{flux} of the traffic flow; \cite{lighthill1955kinematic,richards1956shock} suggested one of the earliest models based on this idea; the LWR model is a non-linear PDE on the density of vehicles.  To overcome the limitations of this model, \cite{payne1971model,whitham2011linear} added a momentum term to the LWR model, which allowed for more complex traffic dynamics than before. However, modeling traffic flow as isotropic~\cite{cassidy1995viethodology, daganzo1995requiem} led to non-physical behavior, such as negative velocity. Based on the observation that traffic flow is anisotropic, \cite{aw2000resurrection,zhang2002non} modified the momentum term in the previous model.  The ARZ model has been used to visually simulate the traffic flow~\cite{sewall2010continuum}, and to solve traffic congestion problems~\cite{yu2019traffic}. Because of the basic assumption, the computational cost of these macroscopic models is proportional to the length of the simulated road, not the number of vehicles therein. Therefore, they are often more computationally efficient than the other models for large-scale environments. However, they are not suitable for simulations where fine details, such as the behavior of individual vehicles, are needed. 

Microscopic models describe the motion of individual vehicles, typically tracking velocity, the bumper-to-bumper distance to its leading vehicle, and the relative velocity between them~\cite{kesting2007general}. The research history dates back to~\cite{gazis1959car, gazis1961nonlinear} and~\cite{newell1961nonlinear}. Modern variations of the model often include behavioral traits of individual vehicles, which make the simulation more realistic and descriptive. \cite{gipps1981behavioural,bando1995dynamical,treiber2000congested,jiang2001full} are such models. These models are widely adopted in current agent-based traffic simulators, such as \cite{lopez2018microscopic}. Since every individual vehicle observes its surrounding environments and decides its actions in these models, we can get more fine-grained details of vehicle motions than in the macroscopic models. However, their computational cost is proportional to the number of vehicles in the scene, which makes them harder to be adopted for large-scale simulations.

Hybrid models seek to combine these models in various fashions to take advantage of their complementary properties; \cite{magne2000towards,bourrel2003mixing,mammar2006highway,sewall2011interactive} fall into this category and show the computational gains we can get from these models. This hybridization is commonly a spatial one, with different parts of a network running under distinct regimes. That is, these models often simulate only the regions of interest with microscopic models for accuracy and use macroscopic models elsewhere to maintain the overall flow correctness while attaining computational efficiency. This typically necessitates a technique for converting between different traffic flow representations~\cite{bourrel2003mixing}. In this work, we adopt this hybrid approach to support large-scale scenarios and let our simulator be more general so that users can select the simulation modes based on their needs.

\subsection{Differentiable models}

Differentiable models have been widely used in graphics and robotics applications like visualization~\cite{david2019mitsuba,li2018diff}, design~\cite{du2020stokes,cascaval2021differentiable}, and control~\cite{heiden2021disect}. This paradigm enables gradient information to flow across complicated functions, and facilitates the machine learning process by improving sample efficiency~\cite{mora2021pods,shen2021high}. They often show better results in finding optimal solutions for various high-dimensional problems~\cite{ma2021diffaqua, Hu2019:ICLR}. 

Related to traffic simulation, there especially have been a number of differentiable simulations for a variety of systems including rigid bodies~\cite{de2018end,qiao2020scalable}, ariticulated bodies~\cite{qiao2021Efficient}, soft bodies~\cite{Geilinger2020add,du2021diffpd,Qiao2021Differentiable}, cloth~\cite{li2022diffcloth}, and fluids~\cite{holl2020phiflow,takahashi2021differentiable}.

Because many traffic models are differentiable in nature, we can apply this technique to traffic simulation and control. Recently, \cite{andelfinger2021differentiable} has presented a differentiable agent-based traffic simulation, and showed how it can be used to solve traffic signal control problems. However, the scope of this work was limited to microscopic models for traffic, and the implementation relies on automatic differentiation, which can be computationally inefficient. In addition to traffic, this hybrid differentiable simulation framework can potentially generalize to other large-scale, multi-agent systems, such as crowds, insects, fluids, grains, etc.~\cite{he2020informative,colas2022interaction,hadrich2021fire,ishiwaka2021foids} as well.

%% file: tex/3_macro.tex
In a macroscopic traffic model, traffic evolution is described by a system of Partial Differential Equations (PDEs) on the road network. 

\subsection{Formulation}
The ARZ~\cite{aw2000resurrection,zhang2002non} model describes vehicle flow in a single lane of traffic through the following system of differential equations in one dimension:

\begin{subequations}
\begin{equation}
\label{eq:arz}
    \mathbf{q}_t + \mathbf{f(q)}_x = 0, \quad \mathbf{f(q)} = \mathbf{q} u =  \begin{bmatrix}
        \rho u \\
        yu
    \end{bmatrix}
\end{equation}
where $\mathbf{q}(x, t) = \left[\rho(x, t), y(x, t)\right]^{\intercal}$; $\rho$ denotes the density of traffic (cars per car length),  $y$ is the relative flow of traffic, and $u$ is the velocity of traffic. 
As per the notation, these terms are observed at a certain position $x$ at a time $t$, and they are related to each other as follows:
\begin{align}
    y(\rho, u) &= \rho\left(u - u_{\mathrm{eq}}(\rho)\right) \\
    u_{\mathrm{eq}}(\rho) &= u_{\max}\left(1 - \rho^{\gamma}\right)
\end{align}
\end{subequations}
where $u_{\max}$ denotes the maximum velocity, or speed limit, of the given lane, and $0 < \gamma < 1$ is a constant tuning parameter. Conceptually, $u_\mathrm{eq}$ represents a `comfortable' velocity based on density. We used $\gamma = 0.5$ in all of the experiments.

Note that the system~\eqref{eq:arz} is a conservative system, where the sum of $\rho$ and $y$ over the entire lane is conserved, excluding the incoming and outgoing fluxes at the borders of the lane.

\subsection{Numerical Solution}
Conservation laws like the ARZ system of PDEs are typically discretized with the Finite Volume Method (FVM)~\cite{leveque2002finite} and integrated explicitly in time. The discretized quantities in a lane are $\mathbf{Q}_{i}^{n}$, where $n$ refers to the time step and $i$ refers to the $i^{\mathrm{th}}$ cell. The solution procedure follows:
\begin{enumerate}
    \item Compute wave speeds $\lambda$ and fluxes $\mathbf{f}$ (from Eq.~\eqref{eq:arz}) by solving the Riemann problem at the interface of each adjacent cell.
    \item Compute global time step $\Delta t \le \max \left|\lambda\right|$; in this work we choose a conservative, constant $\Delta t$.
    \item Integrate: $\mathbf{Q}^{n + 1}_{i} = \mathbf{Q}^{n}_{i} - \frac{\Delta t}{\Delta x}\left[\mathbf{f}(\mathbf{q}(b)) - \mathbf{f}(\mathbf{q}(a))\right]$. $\mathbf{q}(a)$ and $\mathbf{q}(b))$ are the intermediate states identified between $\left(\mathbf{Q}^n_{i-1}, \mathbf{Q}^n_{i}\right)$ and $\left(\mathbf{Q}^n_{i}, \mathbf{Q}^n_{i+1}\right)$ respectively. $\Delta x$ is the length of cell $i$.
\end{enumerate}

Details can be found in~\cite{sewall2011efficient}; Appendix~\ref{appendix-riemann-solution-formulation} reproduces the six relevant cases for the ARZ Riemann problems.

\subsection{Differentiation}
Our differential technique built on this requires analytical gradients of the above; these gradients form the basis of still other gradients that are used in solving more complex traffic control problems.

In our update scheme, we compute the gradients of $\mathbf{Q}^{n+1}_{i}$ using $\mathbf{Q}^{n}_{i-1}$,$\mathbf{Q}^{n}_{i}$, and $\mathbf{Q}^{n}_{i+1}$:
\begin{subequations}
\begin{align}
    \frac{\partial\mathbf{Q}^{n + 1}_{i}}{\partial\mathbf{Q}^{n}_{i - 1}} &= -\frac{\Delta t}{\Delta x}\left[-\mathbf{f}'(\mathbf{q}(a))\frac{\partial \mathbf{q}(a)}{\partial\mathbf{Q}^{n}_{i - 1}}\right]\\
    \frac{\partial\mathbf{Q}^{n + 1}_{i}}{\partial\mathbf{Q}^{n}_{i}} &= \mathbf{I} - \frac{\Delta t}{\Delta x}\left[\mathbf{f}'(\mathbf{q}(b))\frac{\partial \mathbf{q}(b)}{\partial\mathbf{Q}^{n}_{i}} - \mathbf{f}'(\mathbf{q}(a))\frac{\partial \mathbf{q}(a)}{\partial\mathbf{Q}^{n}_{i}}\right] \\
    \frac{\partial\mathbf{Q}^{n + 1}_{i}}{\partial\mathbf{Q}^{n}_{i + 1}} &= -\frac{\Delta t}{\Delta x}\left[\mathbf{f}'(\mathbf{q}(b))\frac{\partial \mathbf{q}(b)}{\partial\mathbf{Q}^{n}_{i + 1}}\right]
\end{align}
\end{subequations}

We thus need the Jacobian $\mathbf{f}(\mathbf{q})$ and the partial derivatives $\frac{\partial \mathbf{q}(a)}{\partial\mathbf{Q}^{n}_{i - 1}}$, $\frac{\partial \mathbf{q}(a)}{\partial\mathbf{Q}^{n}_{i}}$, $\frac{\partial \mathbf{q}(b)}{\partial\mathbf{Q}^{n}_{i}}$, $\frac{\partial \mathbf{q}(b)}{\partial\mathbf{Q}^{n}_{i + 1}}$; this arises from differentiating solutions to Riemann problems as shown in Appendix~\ref{appendix-riemann-solution-differentiation}. The analytical gradients computed here play an important role in accelerating our traffic simulator, as shown in Section~\ref{sec:experiments}.

\subsubsection{Time step sizes} 

We have derived these gradients under the assumption that $\Delta t$ is constant and satisfies the CFL and stability conditions~\cite{leveque2002finite} for the entire simulation. It would be possible to have a dynamic $\Delta t$, since this is also differentiable (the speed $\lambda$ is determined by the eigenvalues of the flux function $\mathbf{f}$).  However, for the sake of the hybrid approach we ultimately use, we use a constant $\Delta t$ to ensure that the microscopic simulation is stable.

\subsubsection{Continuity issues}

We have computed the gradients above assuming that the update scheme is differentiable, and in fact, we can see that it is continuous and piece-wise differentiable across most cases suggested in Appendix~\ref{appendix-riemann-solution}. 

We do not prove it thoroughly here, as it is generally trivial when we keep in mind that $\lim_{u_l \rightarrow u_r} \mathbf{q}_m = \mathbf{q}_l$. However, Case 1 is exceptional; there is a shock between the phase states $q_l$ and $q_m$, which makes the two states discontinuous. Specifically, we cannot guarantee that $q_m$ converges to $q_l$ when $\lambda_s < 0$ converges to $0$. Therefore, the gradients we compute when $\lambda_s$ is near zero could become unstable. However, in most cases in our simulation, these cases rarely happened and did not affect the quality of our solution even when we used the possibly unstable gradients as they are; it is possible that numerical viscosity plays a role here.

%% file: tex/4_micro.tex
In a microscopic traffic model, traffic flow is described by interactions between multiple discrete vehicles that follow certain rules. Here we use the Intelligent Driver Model (IDM)~\cite{treiber2000congested} to simulate such vehicles.

\subsection{Formulation}
Under the microscopic viewpoint, we can describe the state of a discrete vehicle as
 \begin{equation}
    \mathbf{q}(t) = \begin{bmatrix}
        p(t) \\
        v(t)
    \end{bmatrix},
\end{equation}
where $p$ and $v$ denote the position and velocity of the vehicle respectively for a given time $t$. 

Using these states to describe discrete vehicles occupying a lane, IDM describes the motion of each individual vehicle based on its relationship with the vehicle directly ahead of it; the \emph{leading vehicle}. For the $i^\mathrm{th}$ vehicle, let $h(i)^\mathrm{th}$ its leading vehicle. According to the IDM, the acceleration of the $i^\mathrm{th}$ vehicle is determined as follows:
\begin{subequations}
\begin{align}
    {\Delta p} &= p_{h(i)}(t) - p_{i}(t) - {length}_{h(i)} \\
    {\Delta v} &= v_{i}(t) - v_{h(i)}(t) \\
    s_{\mathrm{opt}} &= s_{\min} + v_{i}(t)T_{\mathrm{pref}} + \frac{v_{i}(t){\Delta v}}{2\sqrt{a_{\max}a_{\mathrm{pref}}}} \\
    a_i(t) &= a_{\max}\left[1 - {\left(\frac{v_i(t)}{v_{\mathrm{targ}}}\right)}^{\delta} - {\left(\frac{s_{\mathrm{opt}}}{\Delta p}\right)}^{2}\right]
\end{align}
\end{subequations}

There are various hyperparameters included in the model. They characterize a vehicle's motion; see Appendix~\ref{appendix-idm-hyperparams} for more details. In our experiments, we randomly initialized those hyperparameters for every single vehicle. Note that we compute $s_{\mathrm{opt}}$ before we compute the acceleration term. It represents the \emph{optimal} space that the vehicle should have to avoid collision to its leading vehicle. 

With the acceleration $a_{i}(t)$, we can update the vehicle's state using Euler's method.

\subsection{Differentiation}

Based on this update scheme, we can compute the analytical gradients of the IDM. Since the derivation is straightforward, we offer the analytical gradients in Appendix~\ref{appendix-idm-differentiation}. Note that compared to the macroscopic model, there is no significant theoretical challenge that we have to deal with in the IDM. 

%% file: tex/5_conversion.tex
Our traffic simulation can transition between the continuous, flow-based macroscopic model and the discrete, agent-based microscopic model. We need to consider the differentiation across the interfaces.

\subsection{Macro to Micro}
Given a lane in the macroscopic regime that flows into a microscopic lane, we must convert the continuous representations therein to discrete vehicles at their junction. This can be done in a deterministic or stochastic fashion. 

\subsubsection{Deterministic Instantiation}
Assume that $\rho(t)$ is the density of the traffic flow at the end of a macro lane on time $t$, and $v(t)$ is the velocity. We can set a \emph{flux capacitor} (c.f. \cite{sewall2011efficient}) at the interface; this counts the total number of vehicles $N_d(t)$ that have reached this point:
\begin{equation}
    N_d(t) = \floor*{\int_0^t \rho(t)v(t)dt}
\end{equation}
where $\floor*{x}$ is the floor function that gives the greatest integer less than or equal to the real number $x$. $N_d(t)$ will then start from 0, and we will instantiate a vehicle when it increases by 1. The velocity of the instantiated vehicle is set to be the macroscopic traffic flow's velocity.

\subsubsection{Stochastic Instantiation}
Another possible interpretation of the macro to micro conversion is as a Poisson process. The flux of the traffic flow can be viewed as the intensity of a Poisson process. In each time step of length $dt$, the number of instantiated vehicles $X(t)$ follows a Poisson distribution $X(t) \sim P(\rho(t)v(t)dt)$, $P(X=k)=e^{-\rho(t)v(t)dt}(\rho(t)v(t)dt)^k k!^{-1}$. Similarly, the total number of vehicles $N_s(t)$ can be expressed as
\begin{equation}
    N_s(t) = \int_0^t X(t)
\end{equation}

As above, the velocity of the emitted discrete vehicle is the velocity of the macroscopic state at the junction of the two lanes.

\subsection{Micro to Macro}
Compared to the macro to micro conversion, agent-based information can be converted to a continuum representation more simply. The density of a cell at the interval $(l,r]$ in a macro lane can be defined as
\begin{align}
\label{eq:rho_ia}
    \rho(l, r) = \frac{1}{r-l}\sum_{i=1}^{n} \mathbf{1}_{(l,r]}(p_{i})
\end{align}

\noindent
where $\mathbf{1}(\cdot)$ is the indicator function that identifies vehicles that lie in the interval, $n$ is the number of vehicles and $p_{i}$ is the position of $i^{th}$ vehicle. The velocity of this cell can be the average velocities of all the vehicles therein.

\subsection{Differentiation}
We have described two types of discrete processes: instantiation of vehicles, and the indicator function. Automatic differentiation is able to compute the gradients of velocities $v$ but cannot handle the gradients of densities $\rho$. To backpropagate derivatives to the densities $\rho$, we first create an ancillary variable $a_i = 1$ for each discrete vehicle. We rewrite Equation~\ref{eq:rho_ia} as:
\begin{align}
    \rho(l, r) = \frac{1}{r-l}\sum_{i=1}^{n} \mathbf{1}_{(l,r]}(p_i)\cdot a_i
\end{align}
Since we always set $a_i = 1$, the equation constantly holds for forward simulation. But $a_i$ can also receive the gradients for discrete vehicles from the following macro lanes. Let $L$ be the loss function; then for each vehicle in $(l,r]$:
\begin{equation}
   \PD{L}{a_i}=\PD{L}{\rho(l, r)}
\end{equation}

The next challenge for differentiation is the vehicle instantiation process in the macro to micro conversion. We will start with the deterministic strategy: assume that the total number of vehicles reach $n$ and $n+1$ at time $t_1$ and $t_2$, respectively.  We therefore have the following equation
\begin{equation}
    n+1 = n + \int_{t_1}^{t_2}\rho(t) v(t) dt = n + a_i
\end{equation}
where the $i^{th}$ vehicle is the one instantiated at $t_2$. So for $t\in(t_1,t_2]$, 
\begin{equation}
 \PD{L}{\rho(t)}=\PD{L}{a_i}v(t)dt   
\end{equation}

For the stochastic case, we can first compute the expectation value of the total number of vehicles according to the Poisson distribution
\begin{equation}
E[N_s(t)]=\int_0^tE[X(t)]=\int_0^t\rho(t)v(t)dt.
\end{equation}
We notice that the expectation value of $N_s(t)$ is the same as the flux capacitor in the deterministic strategy. Similarly, we can find time $t_1$ and $t_2$ when the total number of vehicles reach $n$ and $n+1$. For all time step $t\in(t_1,t_2]$, $\PD{L}{\rho(t)}=\PD{L}{a_i}v(t)dt$. Intuitively, if $\PD{L}{a_i}\le 0$, it means the density is higher than the desired level and we need to decrease the intensity of the Poisson process. We will also validate in our numerical experiments that such estimation of gradients can effectively optimize our objective function.

%% file: tex/6_experiments.tex
We show the effectiveness of our differentiable hybrid traffic simulator with application to solve a variety of traffic problems. We first show that we can accelerate the differentiable traffic simulator with the analytical gradients that we have computed above by comparing it against a baseline simulator using automatic differentiation. At the same time, we will justify the use of the hybrid model by comparing its computation time with other approaches in large-scale scenarios. Then, we prove the correctness and efficacy of our analytical gradients by solving parameter estimation problems. Lastly, we conduct experiments with traffic control problems to illustrate how we can integrate our simulator with neural networks and how we can use our simulator to solve real-world traffic problems.

We have implemented our traffic simulator with Python, and used PyTorch 1.9~\cite{paszke2019pytorch} for automatic differentiation. All experiments were run on an Intel\textregistered~Xeon\textregistered~W-2255 CPU @ 3.70GHz, and traffic rendering was mainly based on our own implementation on Unity Engine.


\input{tex/6A_exp_acceleration}

\input{tex/6B_exp_parameter_estimation}

\input{tex/6C_exp_control}

%% file: tex/6A_exp_acceleration.tex
\subsection{Acceleration}

\subsubsection{Analytical Gradients}

Here we present how our analytical gradients can contribute to the acceleration of both forward simulation and backward propagation in our simulator. We have compared our simulator to the baseline simulator that relies on automatic differentiation to compute the gradients in the system. For all of the simulations, we have computed the gradients of the states at the last time step with respect to the states at the initial time step. For macroscopic and microscopic simulations, we used a single lane. For hybrid simulation, we used 3 lanes, which is the minimum case to cover both macro-to-micro and micro-to-macro conversion. The first and third lanes are set as macroscopic lanes and the second one is a microscopic lane. To show how the computation time changes with the scale of the simulation, we have experimented with different scales of settings. 

Table~\ref{speed-comparison} shows that our analytical gradient allows us to run both forward simulation (FW) and back propagation (BP) of the system much faster than the baseline framework. In macroscopic and hybrid simulations, the speedup was up to 4x times for forward simulation, and 18x for back propagation. In the microscopic simulation, the speedup was up to 6x times for forward simulation, and 7x times for back propagation. This shows the computational efficiency of our framework, which is necessary to implement large-scale traffic simulation.

\subsubsection{Hybrid Model}

\begin{figure}
    \centering
    \begin{tikzpicture}
    \begin{axis} [ybar,
                ylabel={\ Frame Per Second (FPS)},
                xlabel={\ $\epsilon$ = Ratio of microscopic vehicles},
                symbolic x coords = {0.01, 0.02, 0.05, 0.1, 0.2, 0.5},
                xtick=data,],
    \addplot coordinates {
        (0.01, 3.57) 
        (0.02, 3.28) 
        (0.05, 2.59) 
        (0.1, 2.08) 
        (0.2, 1.35) 
        (0.5, 0.70)
    };
    \end{axis}
    \end{tikzpicture}
\caption{Comparison on the average frame rate (FPS) under the hybrid model, with different $\epsilon$. The total number of vehicles is set to be 10K, and the $\epsilon$ value denotes the ratio of microscopic vehicles among them.}
\label{speed-comparison-hybrid}
\end{figure}
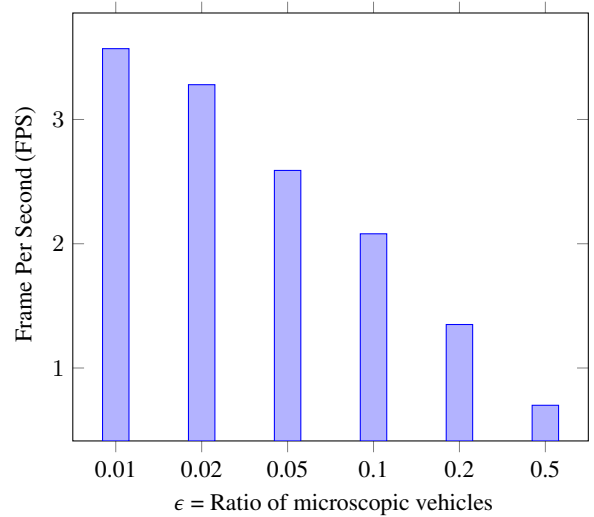

In this paper, we selected the hybrid model to run large-scale traffic simulations in interactive time, while not losing their fidelity. To justify this design choice, here we present the average frame rate for simulating a large environment, which includes approximately 10K vehicles. Note that we can specify the ratio $\epsilon$ of the vehicles which would be simulated using a microscopic model; $\epsilon$ could be regarded as an interpolant between the two models, where $\epsilon = 0$ means macroscopic and $\epsilon = 1$ means microscopic approach. We measured the average frame rate for different $\epsilon$ using single core; see Figure~\ref{speed-comparison-hybrid}. Note that the average frame rates decreases as $\epsilon$ increases and thus the simulation becomes more "microscopic".

\begin{table}[ht]
\caption{Comparison of our technique against automatic differentiation for various types of simulations. Scale denotes number of time steps taken and the number of cells (for macroscopic, and hybrid)/ number of vehicles (microscopic).
FW: forward simulation; BP: backpropagation.  Ours outperforms autodifferentiation in both forward passes and backpropagation, by up to a factor of {\bf 18.77, 7.84, and 15.27} in {\em macroscopic, microscopic}, and {\em hybrid} traffic simulation, respectively. \\}
\label{speed-comparison}
\centering 
\begin{tabular}{c|lrr}
& Scale & 10/1K & 50/5K \\
\hline
\multirow{6}{1.1em}{\rotatebox{90}{Macroscopic}} & FW(Auto) & 5.28s$\pm$0.16s & 125.88s$\pm$5.66s \\
& FW(Ours) & 1.42s$\pm$0.04s & 30.06s$\pm$0.88s \\
& Speedup & 3.71x           & 4.19x            \\
\cline{2-4}
& BP(Auto) & 2.42s$\pm$0.20s & 71.08s$\pm$6.01s \\
& BP(Ours) & 0.18s$\pm$0.03s & 3.79s$\pm$0.25s  \\
& Speedup & {\bf 13.73x}          & {\bf 18.77x}            \\
\hline
\multirow{6}{1.1em}{\rotatebox{90}{Microscopic}}& FW(Auto) & 1.54s$\pm$0.08s & 66.74s$\pm$1.67s \\
& FW(Ours) & 0.45s$\pm$0.02s & 10.86s$\pm$0.24s \\
& Speedup. & 3.42x           & 6.15x            \\
\cline{2-4}
& BP(Auto) & 1.58s$\pm$0.06s & 45.36s$\pm$0.95s \\
& BP(Ours) & 0.20s$\pm$0.03s & 7.10s$\pm$0.42s  \\
& Speedup & {\bf 7.84x}           & {\bf 6.39x}            \\
\hline
\multirow{6}{1.1em}{\rotatebox{90}{Hybrid}}& FW(Auto) & 11.81s$\pm$0.42s & 310.67s$\pm$15.02s \\
& FW(Ours) & 3.41s$\pm$0.06s & 66.52s$\pm$0.47s \\
& Speedup & 3.47x           & 4.67x            \\
\cline{2-4}
& BP(Auto) & 2.70s$\pm$0.11s & 111.06s$\pm$3.09s \\
& BP(Ours) & 0.38s$\pm$0.02s & 7.27s$\pm$0.13s  \\
& Speedup & {\bf 7.13x}          & {\bf 15.27x}            \\
\hline
\end{tabular}
\vspace*{1em}
\end{table}


%% file: tex/6B_exp_parameter_estimation.tex
\begin{figure}[ht]
\vspace*{-1.5em}
\begin{center}
\begin{tabular}{@{}c@{\hspace{1mm}}c@{\hspace{1mm}}@{}}
    \subfloat[Macroscopic]{
        \includegraphics[width=0.5\linewidth]{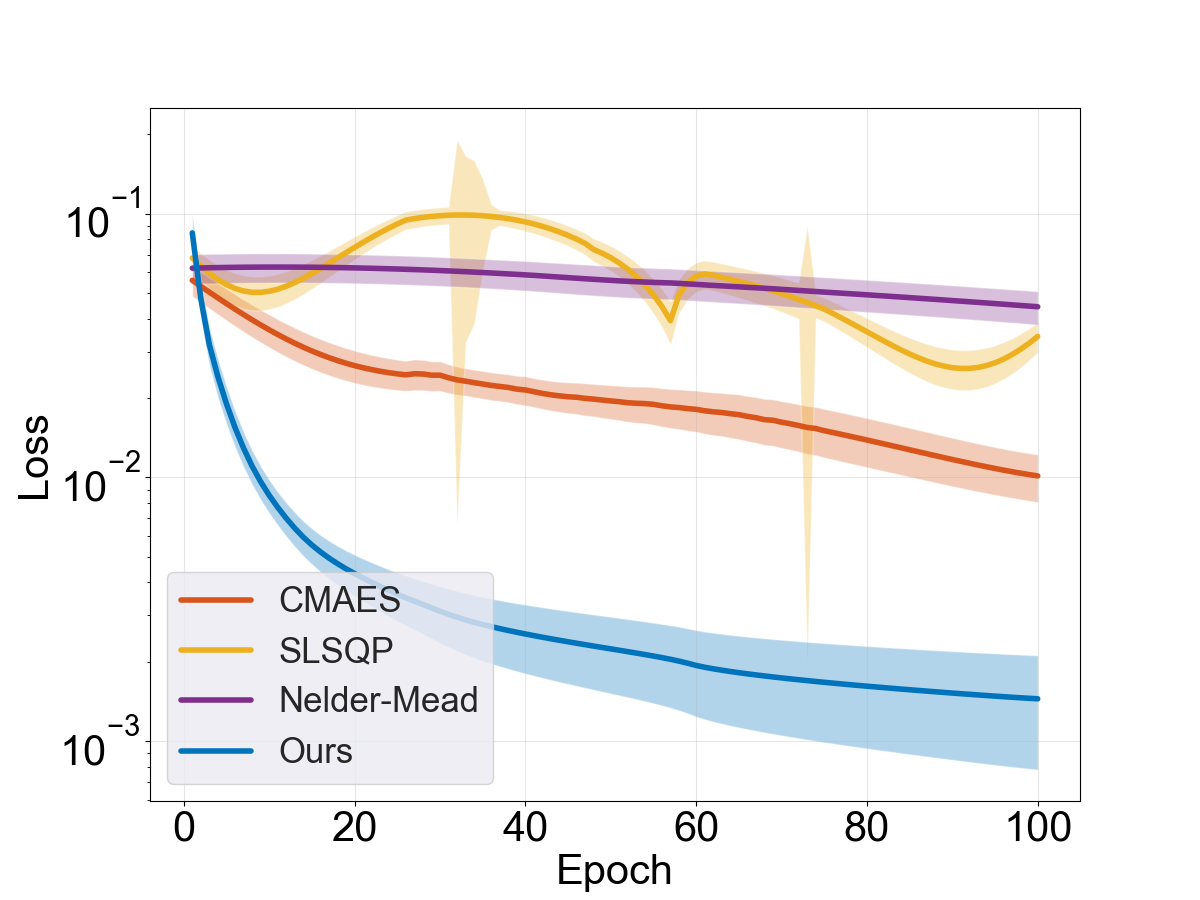}
    } &
    \subfloat[Microscopic]{
        \includegraphics[width=0.5\linewidth]{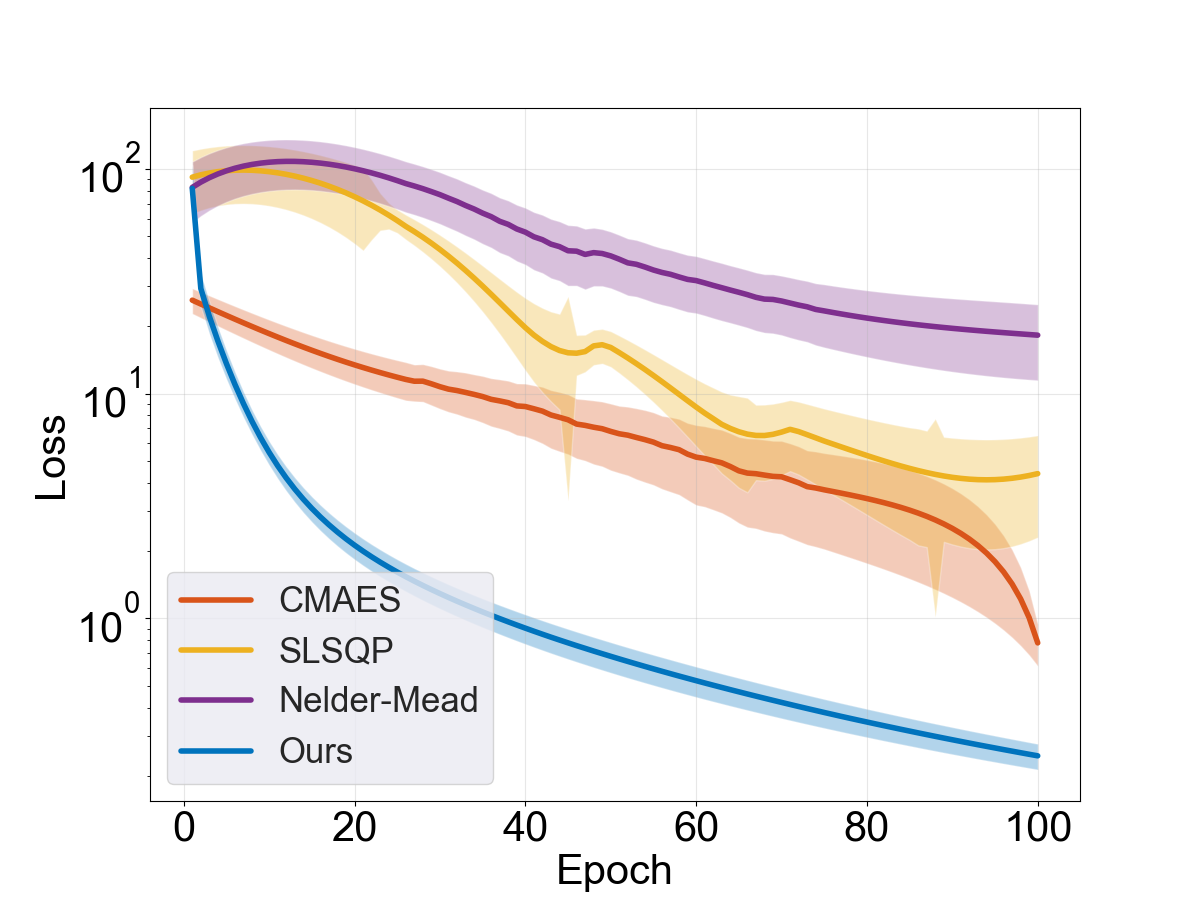}
    } \\
    \subfloat[Hybrid (Macro-Micro-Macro)]{
        \includegraphics[width=0.5\linewidth]{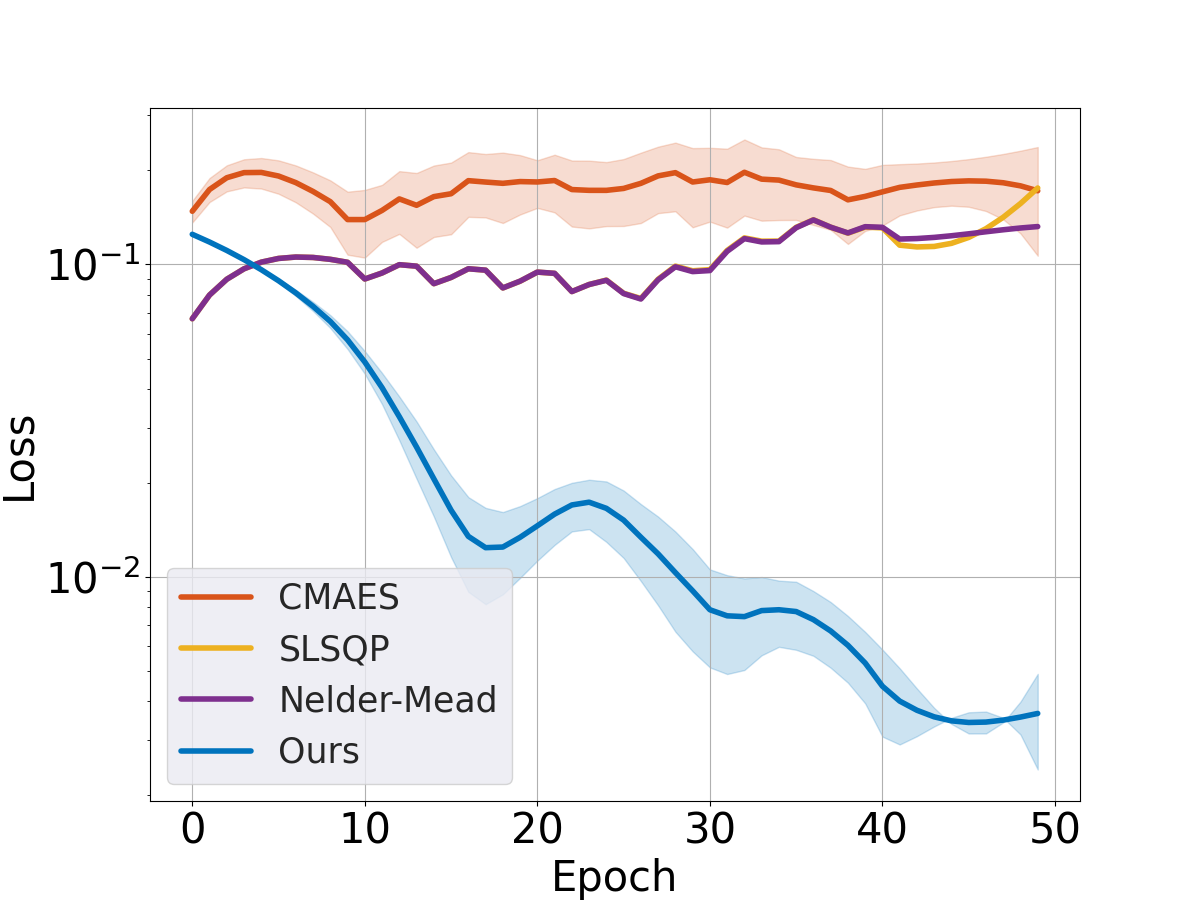}
    } &
    \subfloat[Hybrid (Micro-Macro-Micro)]{
        \includegraphics[width=0.5\linewidth]{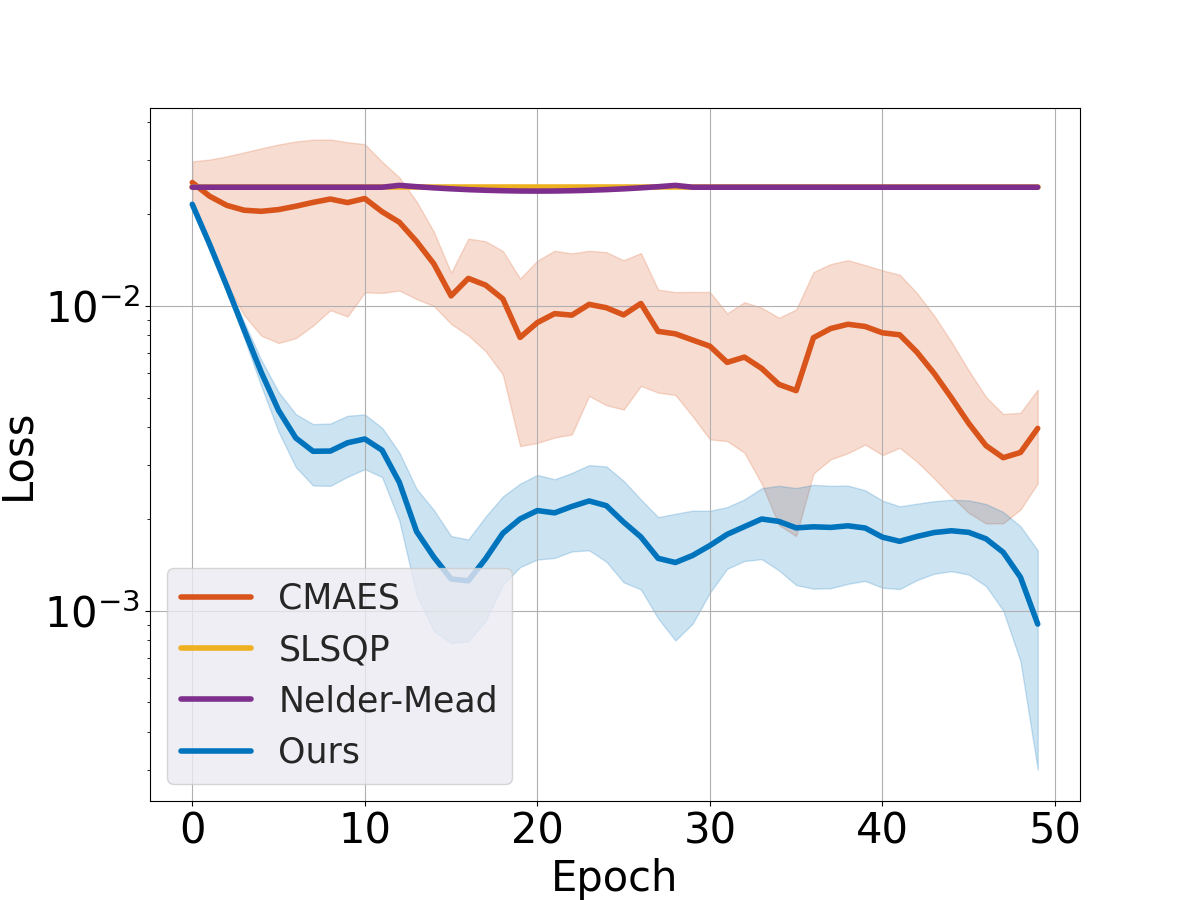}
    }
    \\
\end{tabular}
\caption{{\bf Parameter estimation problem.} For given final states of the system, one has to estimate the initial states that lead to the final states. (a)--(d) Our results rendered in blue exhibit much faster convergence to the correct solutions than the other state-of-the-art gradient-free optimization methods by up to {\em one order of magnitude}.}
\label{fig-param-est}
\end{center}
\vskip -0.1in
\end{figure}

\subsection{Parameter Estimation Problem}

One of the most fundamental problems that we can solve with our framework is the parameter estimation problem. There are many parameters involved in our system; the speed limit is the most typical example. In our formulation, we focus on retrieving preceding states in the system based on given subsequent ones. The simplest form of this problem would be, assuming $0 \leq t \leq 1$, estimating the initial state $\mathbf{q}(0)$ given the last state $\mathbf{q}(1)$. This problem is considered to be the most basic formulation to solve, as it only requires fundamental gradients computed analytically.

In our experiments, the simulation length was set as 10 seconds for every setting. Theoretically, we can use any simulation lengths for this experiment, but we found out that too long simulation lengths make this problem much harder than it should be and 10 seconds is a proper length to prove our approach's efficacy.  Under this setting, we have to estimate the initial traffic states that would end up in the given final traffic states in 10 seconds. For a proposed estimate $\mathbf{q}_{\mathrm{est}}(0)$, we can compute the estimation error with the following loss function, where $n$ denotes the size of the state vector:
\begin{equation*}
    L = \frac{1}{n}\left|\mathbf{q}(1) - \mathbf{q}_{\mathrm{est}}(1)\right|^2
\end{equation*}

We have compared our framework with other gradient-free optimization algorithms: CMA-ES~\cite{hansen2006cma}, SLSQP~\cite{kraft1988software}, and Nelder-Mead~\cite{nelder1965simplex} algorithm. Experiments were run five times with randomly initialized initial states for each of the algorithms and for each of the simulation modes. In the hybrid setting, we used 3 sequentially connected lanes. One of the experiments was done for macro-micro-macro lanes, and the other was done for micro-macro-micro lanes. The loss graph for each setting is shown in Figure~\ref{fig-param-est}. We can observe that our framework retrieves more precise initial states faster than other algorithms for every setting.


%% file: tex/6C_exp_control.tex
\subsection{Traffic Control Problems}

Our differentiable traffic simulator can be integrated with neural networks to enhance its ability to learn and control a given task. To test our framework's capability, we have experimented with the ``intersection signal control problem'' (ITSCP) for macroscopic and hybrid simulation, and the pace car problem for microscopic simulation. 

For all the experiments, we used two fully connected layers as our neural network, where each of the layers has 256 nodes. The neural network is trained to emit proper control input, which would be fed into our simulator. Then we can compute the gradient of the aggregated reward of each episode with respect to each control input therein, and apply gradient descent to directly maximize the reward. Note that this approach is same as analytical policy gradient (APG) method. Then we compare our results to other baseline gradient-free RL algorithms, as they are widely used to solve these kinds of control problems.

\begin{figure}[t]
\centering
    \includegraphics[width=0.9\linewidth]{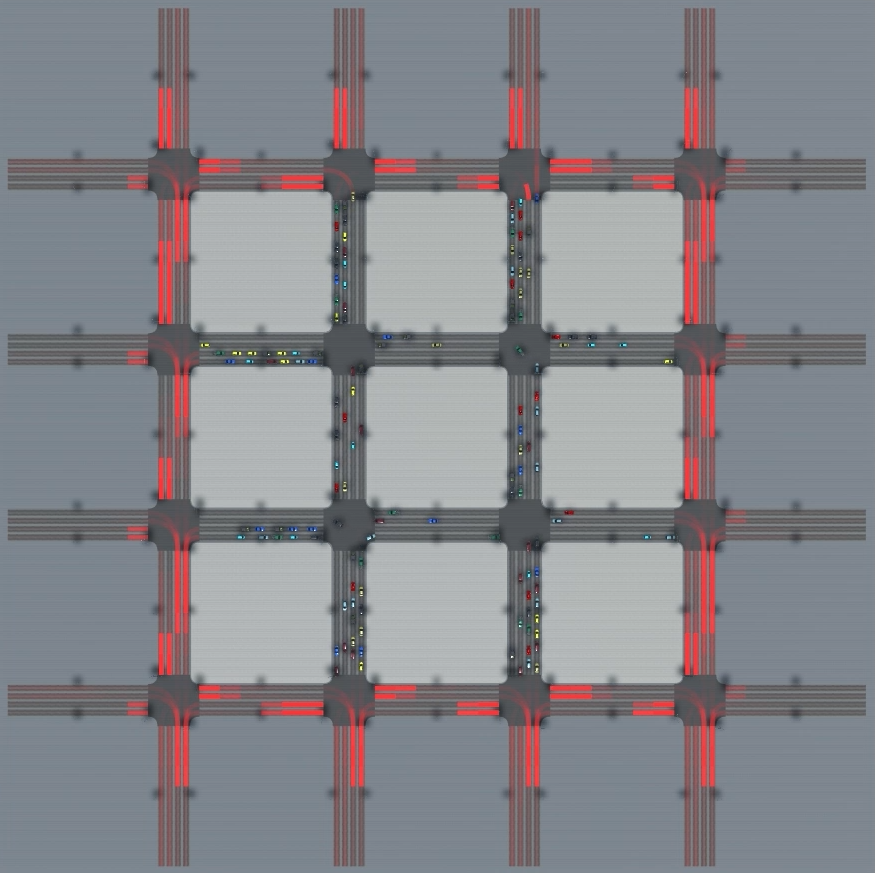}
    \vspace{-1em}
    \caption{{\bf 4x4 Grid of Intersections.} Hybrid traffic simulation at a 4x4 grid of intersections: the density of each cell is rendered in red, which shows the volume of traffic agents to cross the intersection. One has to optimize time allocations for traffic lights at each intersection to maximize overall traffic flow.}
\label{fig:intersection-env}
\vspace*{-1em}
\end{figure}

\begin{figure}[t]
\vspace*{-1.5em}
\centering
\begin{tabular}{@{}c@{\hspace{1mm}}c@{\hspace{1mm}}@{}}
    \subfloat[Macroscopic] {
        \includegraphics[width=0.5\linewidth]{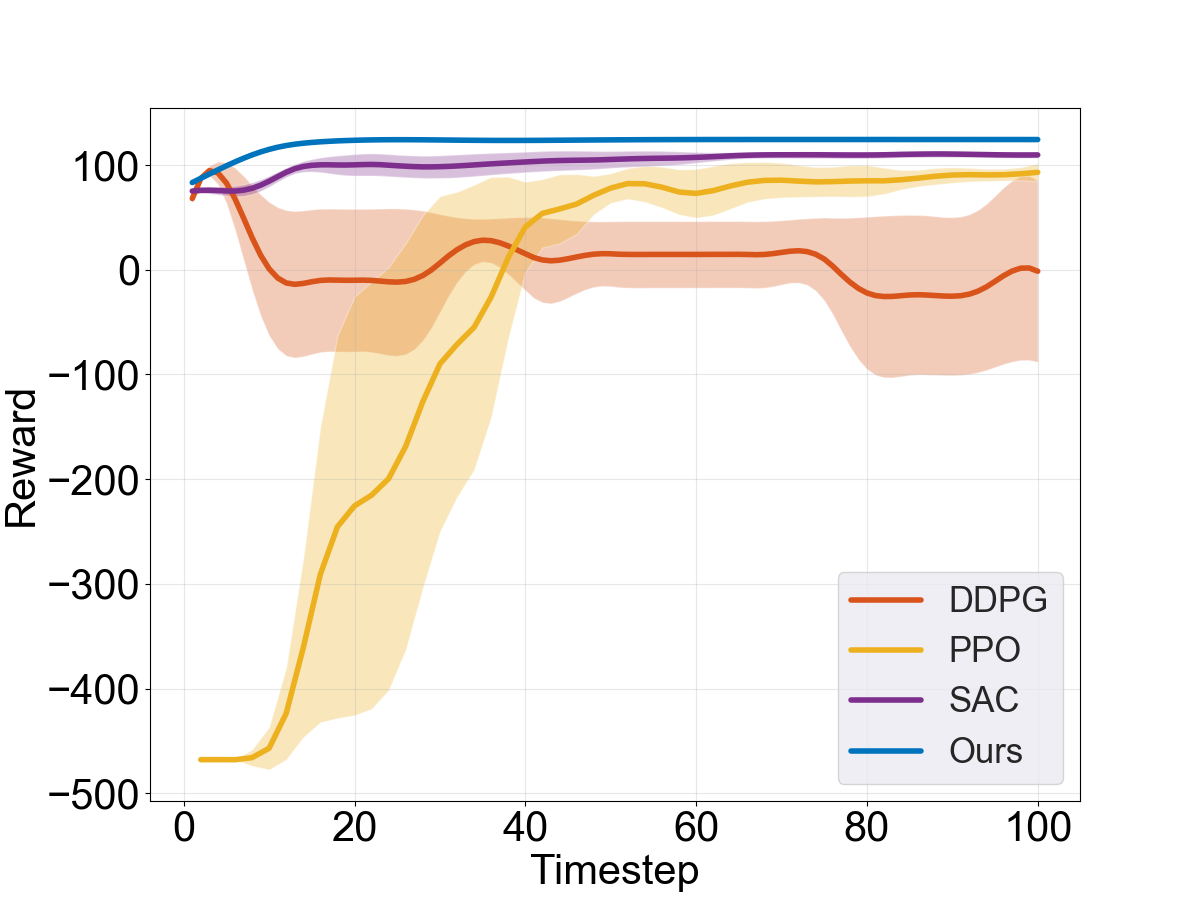}
    } &
    \subfloat[Hybrid] {
        \includegraphics[width=0.5\linewidth]{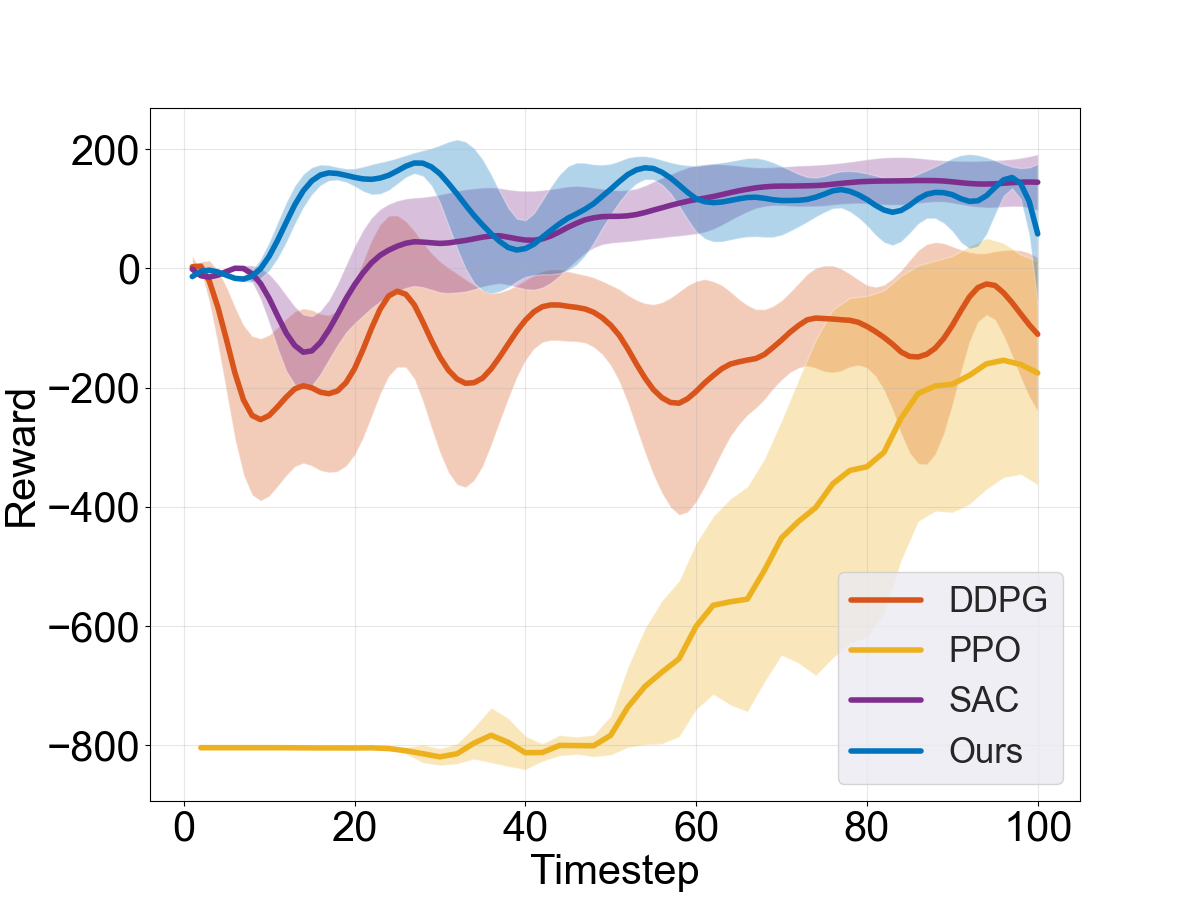}
    } \\
\end{tabular}

\vspace{-1em}
\caption{ {\bf ITSCP.} (a), (b) Learning graphs for macroscopic and hybrid simulations. Our framework shows better convergence rates and results than other RL algorithms.}
\label{fig:itscp-results}
\end{figure}

\begin{figure*}[t]
\vspace{-1.0em}
\centering
    \begin{tabular}{c}
        \subfloat[Time = 0 sec, Initial States]
        {\includegraphics[width=0.95\linewidth]{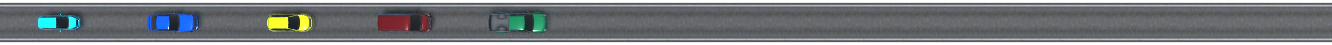} \vspace{-1em}}
        \\
        \subfloat[Time = 10 sec, Target Speed Limit = 30 m/sec]
        {\includegraphics[width=0.95\linewidth]{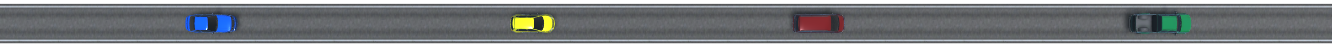} \vspace{-1em}}
        \\
        \subfloat[Time = 20 sec, Target Speed Limit = 10 m/sec]
        {\includegraphics[width=0.95\linewidth]{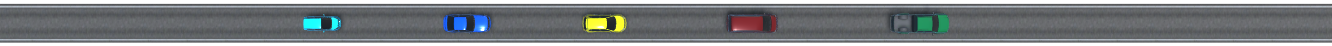} \vspace{-1em}}
        \\
    \end{tabular}
\vspace*{-0.5em}    
\caption{ {\bf Pace Car Problem.} Sequence of simulation images in time order, from top to bottom. (a) A pace car rendered in green guides following vehicles to maintain a target speed limit, which varies over time. It fails when it runs too fast, too slow, or goes out of the lane. (b) When a high target speed limit (30 m/sec) is given, the pace car accelerates fast, and keeps a wide distance from its following vehicles to let them run fast. (c) However, when the target speed limit is reduced (10 m/sec), it swiftly decelerates and keeps short distance from the following vehicles, to prevent them from running over the target speed limit.}
\label{fig:pace-car}
\end{figure*}

\subsubsection{ITSCP}

ITSCP is one of the most widely studied traffic control problems, because of its significance in controlling traffic congestion in urban environment~\cite{eom2020traffic}. We have devised a scenario that falls into this category, where an agent has to compute the optimal time allocations for traffic lights (Figure~\ref{fig:intersection-env}). 

In this scenario, there are two different traffic lights at a given intersection: one is green across the West-East direction (WE-light), and the other one is across the North-South direction (NS-light). When one of the lights is turned on, the other one has turned off automatically. In a single signal phase, the WE-light is turned on first, and then the NS-light is turned on next. There are multiple signal phases during the entire simulation, and the agent has to determine optimal time allocations for each light in each signal phase. 

The input to the neural network is given as the schedule of incoming traffic flow for every lane. Our experiments were inspired by a synthetic traffic scenario devised for another ITSCP experiment~\cite{wei2018intellilight}. We also followed a conventional scheme to compute the reward~\cite{eom2020traffic}. The reward for an action is computed as the weighted sum of traffic flow that crosses the intersection ($R_{f}$) and the length of waiting queue in each lane ($R_{q}$), which is formulated as follows:
\begin{equation*}
    R = c_{1} \cdot R_{f} + c_{2} \cdot R_{q}. 
\end{equation*}
For our experiments, $c_{1}$ is set as 1 and $c_{2}$ is set as -1, to maximize the traffic flow and minimize the length of waiting queue. In macroscopic environments, the length of the queue is computed as the number of cells that have speed less than certain threshold. In hybrid environments, the traffic flow is computed as the number of discrete vehicles crossing the intersection, as the central part of an intersection is simulated with a microscopic model.

For comparison, we used three baseline model-free RL algorithms; DDPG~\cite{lillicrap2015continuous}, PPO~\cite{schulman2017proximal}, and SAC~\cite{haarnoja2018soft}. Note that these algorithms do not use gradient information that our simulator provides. In contrast, our gradient-based optimization algorithm uses this information to optimize the objective function. Experiments were run five times for each algorithm, and Figure~\ref{fig:itscp-results} shows the learning graphs of both macroscopic and hybrid environment. In the macroscopic setting, our framework converged to the near-optimal solution very quickly, while DDPG failed to learn at all and PPO and SAC learned, but not to the extent of ours. In hybrid setting, our framework and SAC both succeeded in converging to the best solution, but ours converged faster than SAC. Also, Table~\ref{control-max-reward-comparison} shows that our best reward is better than any other algorithms for both settings.

\begin{figure}[t]
\vspace{-1.0em}
\centering
    \begin{tabular}{c}
        {\includegraphics[width=1.0\linewidth]{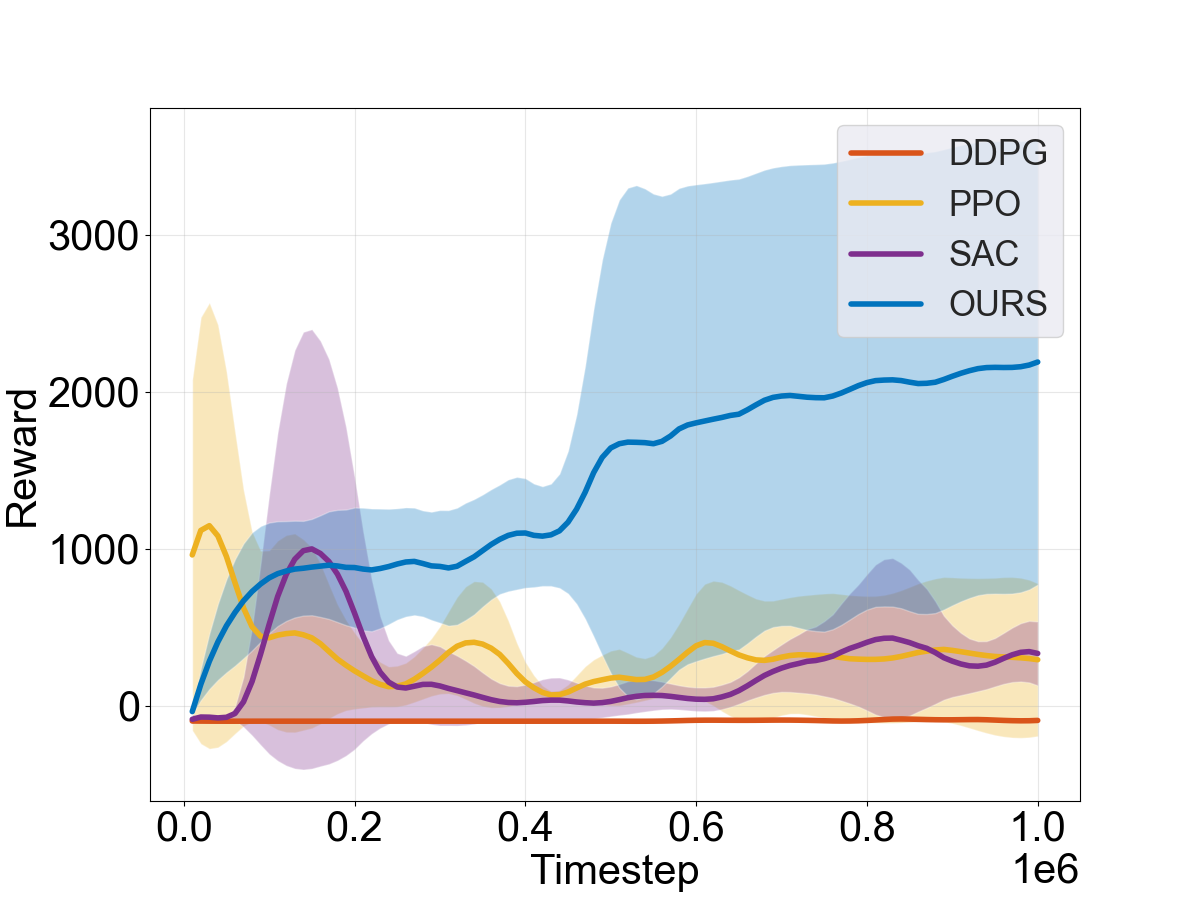}}
    \end{tabular}
\vspace*{-0.5em}    
\caption{ {\bf Pace Car Learning Graph.} Our result rendered in blue shows far better performance (as indicated by higher reward values) than other algorithms, which make no significant gains in learning even after long periods.}
\label{fig:pace-car-result}
\end{figure}

\begin{table}[t]
\caption{Comparison of maximum rewards from each control problem. Our framework is able to find better solutions (indicated by higher reward values) than the other baseline, model-free RL algorithms (DDPG, PPO, or SAC) for every experiment setting (Macro, Micro, or Hybrid).}
\label{control-max-reward-comparison}
\centering 
\begin{tabular}{c|lrrr}
& Ours & DDPG & PPO & SAC \\
\hline
Macro & \textbf{124.38} & 113.62 & 104.53 & 115.73 \\
\hline
Micro & \textbf{4482.87} & 4.97 & 3218.19 & 3373.99 \\
\hline
Hybrid & \textbf{208.48} & 174.18 & 158.18 & 191.51 \\
\hline
\end{tabular}
\vspace*{-0.5em}
\end{table}

\subsubsection{Pace Car Problem}

In this problem, we assume there is a discrete pace car, which runs in front of other vehicles that follow the IDM. The pace car leads the vehicles to enforce a specified target speed limit, which is often smaller than the lane's speed limit (Figure~\ref{fig:pace-car}).

The optimal solution for this problem is clear to the human observer; all vehicles should maintain the desired speed limit. However, the pace car has to determine both acceleration and steering at every frame to remain on the road and lead the vehicles at the same time. To introduce even more complexity, in our experiments, we used a speed limit target that varies over time. Therefore, the neural network receives other vehicle's states and the speed limit as input, and has to compute the optimal acceleration and steering for next several frames, which is quite challenging. The reward is formulated as follows:
\begin{equation}
    R = \sum_{i=1}^{t}\sum_{j=1}^{n}C_{\max} - (v_{\mathrm{targ}} - v_{i, j})^{2}
\end{equation}
where $C_{\max}$ denotes an arbitrary constant value, $t$ and $n$ denote number of frames and number of vehicles respectively. $v_{\mathrm{targ}}$ and $v_{i, j}$ represent the target speed limit and $j^\mathrm{th}$ vehicle's speed at the given frame $i$.

We also used three baseline RL algorithms for comparison, each of which were run 5 times. The learning graph in Figure~\ref{fig:pace-car-result} and the maximum rewards in Table~\ref{control-max-reward-comparison} show that our framework achieved far better overall reward than the other algorithms. PPO and SAC also succeeded in accomplishing initial high rewards but their performance soon plunged. Our framework  continuously learned and achieved better and  improved results.

%% file: tex/conclusions.tex
\begin{figure}[ht]
\vspace{-0.5em}
\centering
\includegraphics[width=1.0\linewidth]{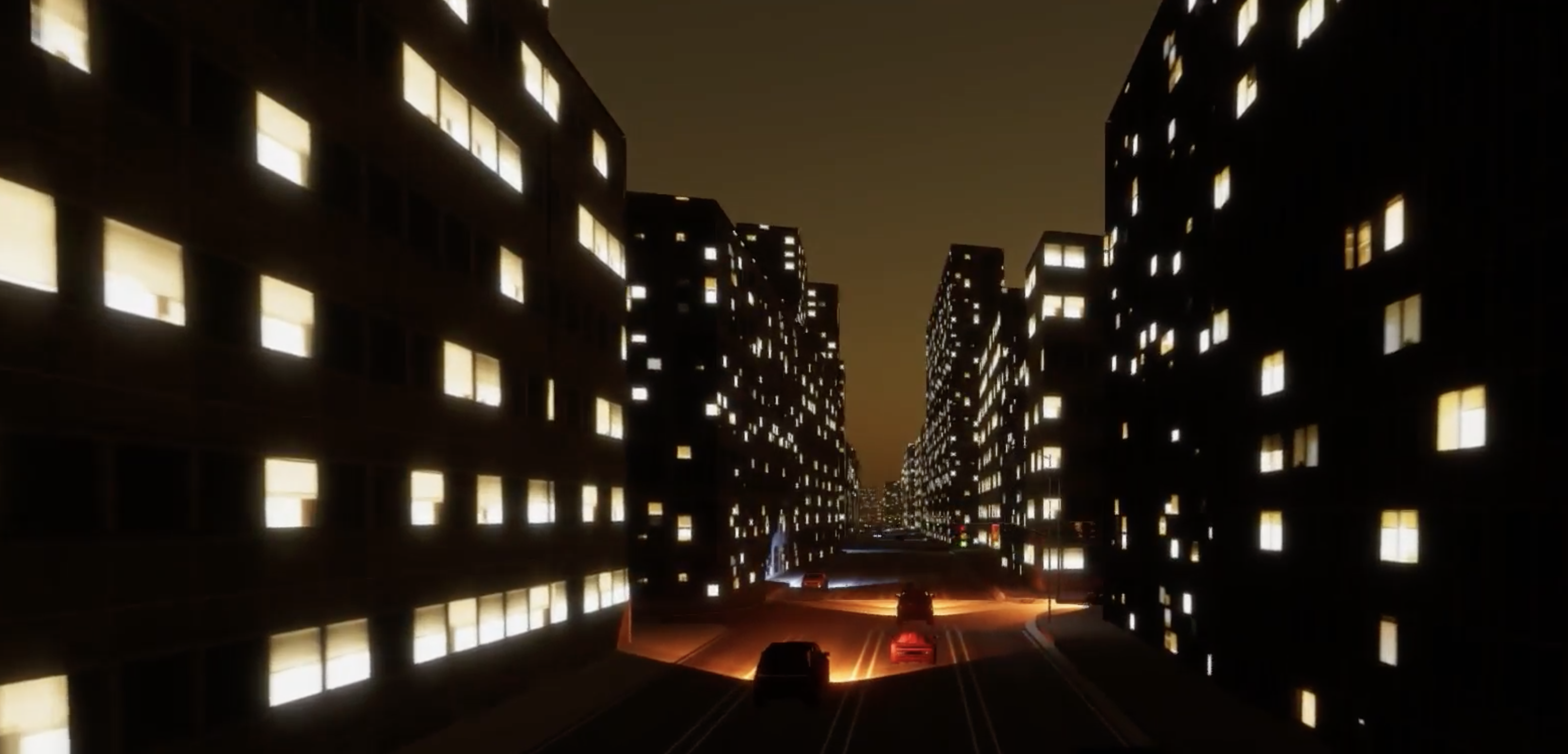} \\
\includegraphics[width=1.0\linewidth]{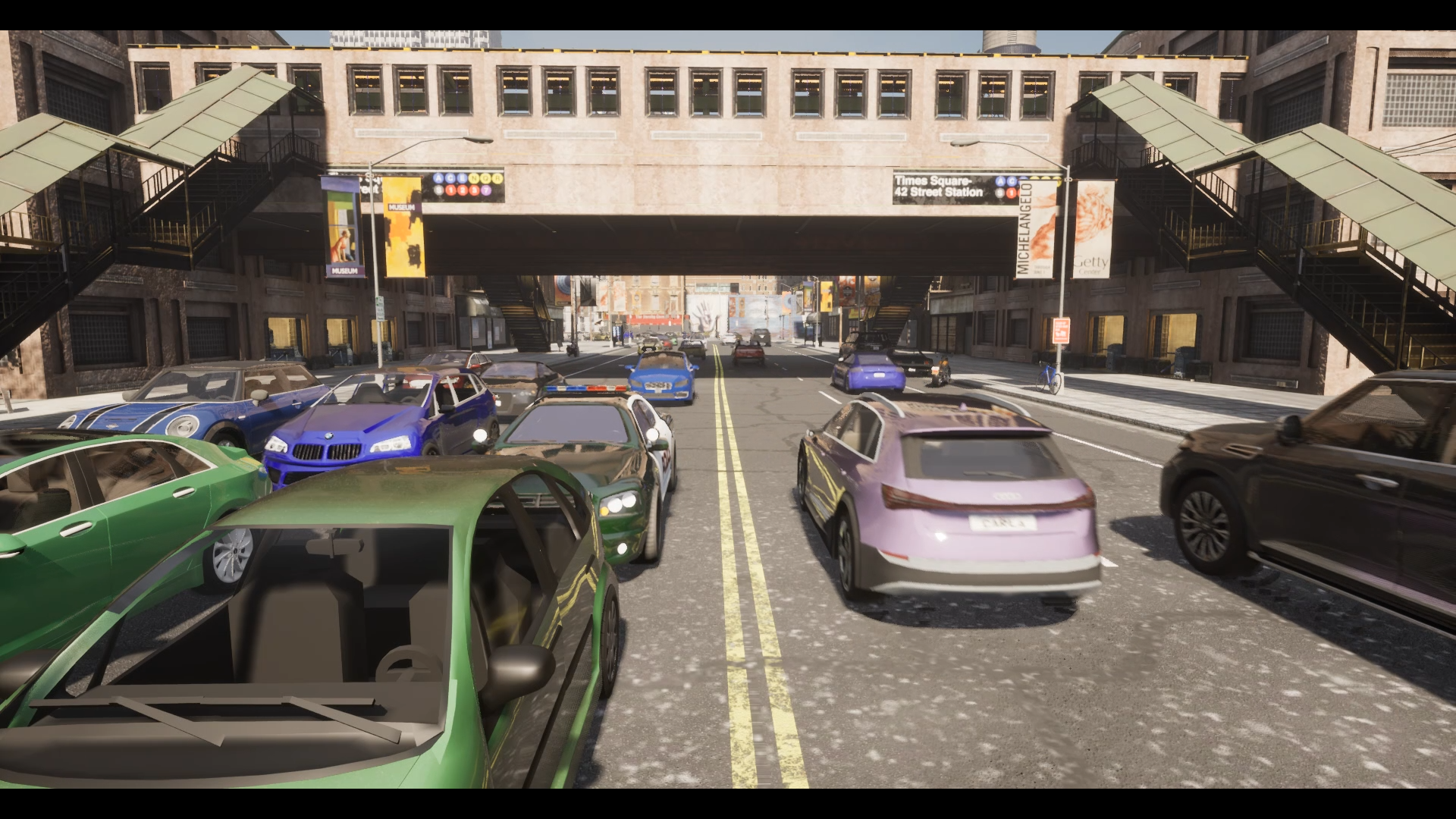} \\
\includegraphics[width=1.0\linewidth]{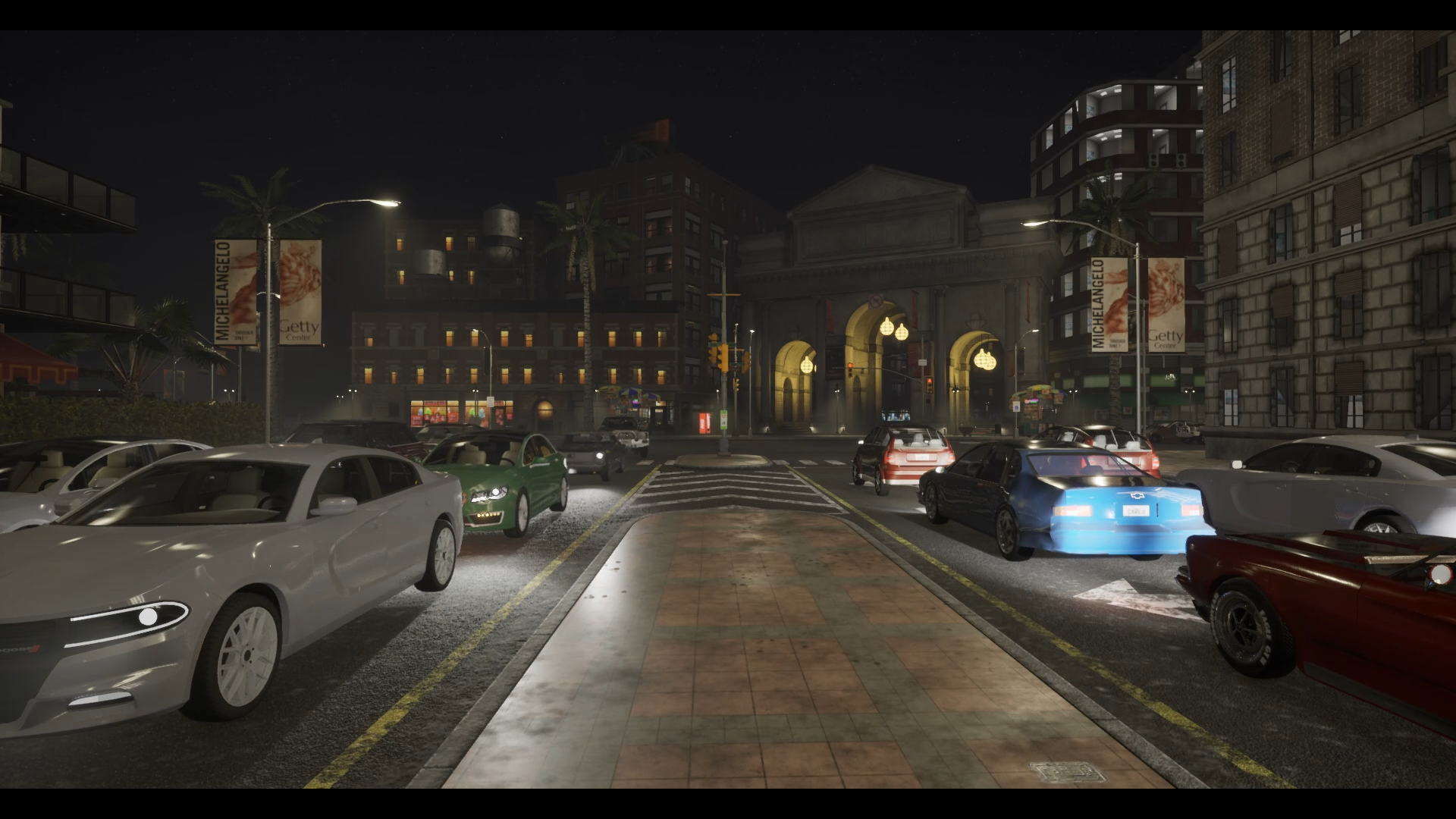}
\vspace{-2em}
\caption{ {\bf Large-scale Hybrid Traffic simulation.} These traffic scenes consist of many hundreds of vehicles across an urban scene simulated and controlled in real time using a differentiable hybrid traffic model.}
\vspace{-0.5em}
\label{fig:real-time}
\end{figure}

We have proposed a novel differentiable hybrid traffic simulator. In our simulator, gradients of traffic states across the time steps are computed analytically and propagated across lanes modeled under complimentary but deeply different regimes. This was made possible with our novel discrete-continuous differentiable conversion module. Further, our analytic formulation is much more efficient than one based on automatic differentiation; therefore, our technique can offer real-time traffic simulation, as shown in Figure~\ref{fig:real-time}, that would not have been possible otherwise.

We have also shown that we can use the gradients to solve classic traffic problems. For the parameter estimation problems, our framework is able to find far better estimates than other gradient-free optimization algorithms. For control problems, our framework succeeded in finding near-optimal policies, which was not possible with other model-free RL algorithms.

\subsection{Generalization} 
The concept of hybrid simulation is applicable to many different types of dynamical systems, such as fluids~\cite{mohamed2010review,golas2012large,narain2010free}, crowd~\cite{treuille2006continuum,narain2009aggregate}, and many other multi-agent systems~\cite{zheng2017consensus}. 
And, our method to differentiate the conversion between macro and micro traffic flow is applicable to a broad range of dynamical systems that are simulated by hybrid techniques, transitioning between continuous and discrete simulation domains. 
In general, other deterministic or stochastic instantiation process can also take advantage of our technique introduced in this paper. For example, the micro and macro representations for crowd share many similarities with traffic. The conversion between individual pedestrians and the crowd flow is analogous to the process of instantiating and removing vehicles in the roads.
We also hypothesize that the algorithmic and computational framework on differentiable hybrid simulation, as proposed here, can also be extended to hybrid control of complex and autonomous systems~\cite{branicky1998unified,fierro2001hybrid}.

\subsection{Limitations and Future Directions}
Given these promising results, we plan to further augment the capability and application of our simulator. 
First, while hybrid techniques are naturally suited to metropolitan-scale scenarios, we have not yet applied this differentiable hybrid technique to these; we expect to achieve even higher performance improvements from these mega-scale experiments. We can also integrate realistic vehicle motions, such as lane changing, to further capture complex traffic dynamics in higher fidelity. Finally, we used simulated data for solving traffic control problems in this paper, but we expect that our simulator is applicable to real-world data as well.

\begin{acks}
This research is supported in part by Dr. Barry Mersky and Capital One E-Nnovate Endowed Professorships, and ARL Cooperative Agreement W911NF2120076.
\end{acks}

%% file: tex/appendix_riemann_solution.tex
\subsection{Formulation}
\label{appendix-riemann-solution-formulation}

For the given left state $\mathbf{q_l}$ and right state $\mathbf{q_r}$, we can determine the intermediate state $\mathbf{q_0}$ as follows.

\mypara{Case 0 : $u_l = u_r$}
\begin{equation}
    q_0 = q_l
\end{equation}

\mypara{Case 1 : $u_l > u_r$}
\begin{equation}
    q_0 = \begin{cases}
        q_l & \mbox{if } \lambda_s \geq 0 \\
        q_m & \mbox{if } \lambda_s < 0 \\
    \end{cases}
    (\lambda_s = \frac{\rho_m u_m - \rho_l u_l}{\rho_m - \rho_l})
\end{equation}

\mypara{Case 2 : $u_r - u_{max}\rho_l^{\gamma} < u_l < u_r$}
\begin{equation}
    q_0 = \begin{cases}
        q_l & \mbox{if } \lambda_{0l} \geq 0 \\
        q_m & \mbox{if } \lambda_{0m} \leq 0 \\
        \tilde{q}(0) & \mbox{if } \lambda_{0l} < 0, \lambda_{0m} > 0
    \end{cases}
\end{equation}
where
\begin{align*}
    \lambda_{0l} &= u_l - u_{max}\gamma\rho_l^{\gamma} \\
    \lambda_{0m} &= u_m - u_{max}\gamma\rho_m^{\gamma} \\
                &= u_r - u_{max}\gamma\rho_l^{\gamma} + \gamma(u_r - u_l).
\end{align*}

\mypara{Case 3 : $u_l \leq u_r - u_{max}\rho_l^{\gamma}$}
\begin{equation}
    q_0 = \begin{cases}
        q_l & \mbox{if } \lambda_{0l} \geq 0 \\
        \tilde{q}(0) & \mbox{if } \lambda_{0l} < 0
    \end{cases}
\end{equation}

\mypara{Case 4 : $\rho_l = 0, \rho_r > 0$}
\begin{equation}
    q_0 = q_l = \begin{bmatrix}
        0 \\
        0
    \end{bmatrix}
\end{equation}

\mypara{Case 5 : $\rho_l > 0, \rho_r = 0$}
\begin{center}
    Same as Case 3.
\end{center}

From above, $q_m$ is given as
\begin{align*}
    \rho_m &= {(\rho_l^{\gamma} + \frac{u_l - u_r}{u_{max}})}^{\frac{1}{\gamma}} \\
    u_m &= u_r \\
    y_m &= \rho_m(u_m - u_{eq}(\rho_m)),
\end{align*}
and $\tilde{q}(0)$ is given as
\begin{align*}
    \tilde{\rho}(0) &= (\frac{u_l + u_{max}\rho_l^{\gamma}}{(\gamma + 1)u_{max}})^{\frac{1}{\gamma}} \\
    \tilde{u}(0) &= \frac{\rho_l}{\rho_l + 1}(u_l + u_{max}\rho_l^{\gamma}) \\
    \tilde{y}(0) &= \tilde{\rho}(0)(\tilde{u}(0) - u_{eq}(\tilde{\rho}(0))).
\end{align*}

\subsection{Differentiation}
\label{appendix-riemann-solution-differentiation}

We can see from the above formulation that we have to compute the partial derivatives of $q_l, q_m,$ and $\tilde{q}(0)$ with respect to $q_l$ and $q_r$, as those values comprise the solutions of the Riemann problem. We can compute the partial derivatives as shown below.

The Jacobian of $\mathbf{f}$ is:
\begin{equation}
    \mathbf{f}'(\mathbf{q}) = \begin{bmatrix}
        u_{eq} + \rho u_{eq}' & 1 \\
        yu_{eq}' - \frac{y^2}{\rho ^2} & \frac{2y}{\rho} + u_{eq}
    \end{bmatrix}
\end{equation}

\mypara{Partial derivatives of $\mathbf{q_l}$}
\begin{align*}
    \frac{\partial q_l}{\partial q_l} &= I_{2 \times 2}
\end{align*}
\begin{align*}
    \frac{\partial q_l}{\partial q_r} &= \begin{bmatrix}
        0 & 0 \\
        0 & 0
    \end{bmatrix}
\end{align*}

\mypara{Partial derivatives of $\mathbf{q_m}$}
\begin{align*}
    \frac{\partial q_m}{\partial q_l} &= \begin{bmatrix}
        \frac{\partial \rho_m}{\partial \rho_l} & \frac{\partial \rho_m}{\partial y_l} \\
        \frac{\partial y_m}{\partial \rho_l} & \frac{\partial y_m}{\partial y_l} \\
    \end{bmatrix}
\end{align*}
where
\begin{align*}
    \frac{\partial \rho_m}{\partial \rho_l} &= \frac{1}{\gamma}(\rho_l^{\gamma} + \frac{u_l - u_r}{u_{max}})^{\frac{1 - \gamma}{\gamma}}(\gamma \rho_l^{\gamma - 1} + \frac{1}{u_{max}}\frac{\partial u_l}{\partial \rho_l}), \\
    \frac{\partial \rho_m}{\partial y_l} &= \frac{1}{\gamma}(\rho_l^{\gamma} + \frac{u_l - u_r}{u_{max}})^{\frac{1 - \gamma}{\gamma}}(\frac{1}{u_{max}}\frac{\partial u_l}{\partial y_l}), \\
    \frac{\partial y_m}{\partial \rho_l} &= \frac{\partial \rho_m}{\rho_l}(u_m - u_{eq}(\rho_m)) + \rho_m(-u'_{eq}(\rho_m)\frac{\partial \rho_m}{\partial \rho_l}), \\
    \frac{\partial y_m}{\partial y_l} &= \frac{\partial \rho_m}{y_l}(u_m - u_{eq}(\rho_m)) + \rho_m(-u'_{eq}(\rho_m)\frac{\partial \rho_m}{\partial y_l}). \\
\end{align*}

\begin{align*}
    \frac{\partial q_m}{\partial q_r} &= \begin{bmatrix}
        \frac{\partial \rho_m}{\partial \rho_r} & \frac{\partial \rho_m}{\partial y_r} \\
        \frac{\partial y_m}{\partial \rho_r} & \frac{\partial y_m}{\partial y_r} \\
    \end{bmatrix}
\end{align*}
where
\begin{align*}
    \frac{\partial \rho_m}{\partial \rho_r} &= \frac{1}{\gamma}(\rho_l^{\gamma} + \frac{u_l - u_r}{u_{max}})^{\frac{1 - \gamma}{\gamma}}(-\frac{1}{u_{max}}\frac{\partial u_r}{\partial \rho_r}), \\
    \frac{\partial \rho_m}{\partial y_r} &= \frac{1}{\gamma}(\rho_l^{\gamma} + \frac{u_l - u_r}{u_{max}})^{\frac{1 - \gamma}{\gamma}}(-\frac{1}{u_{max}}\frac{\partial u_r}{\partial y_r}), \\
    \frac{\partial y_m}{\partial \rho_r} &= \frac{\partial \rho_m}{\rho_r}(u_m - u_{eq}(\rho_m)) + \rho_m(\frac{\partial u_r}{\partial \rho_r} - u'_{eq}(\rho_m)\frac{\partial \rho_m}{\partial \rho_r}), \\
    \frac{\partial y_m}{\partial y_r} &= \frac{\partial \rho_m}{y_r}(u_m - u_{eq}(\rho_m)) + \rho_m(\frac{\partial u_r}{\partial y_r} - u'_{eq}(\rho_m)\frac{\partial \rho_m}{\partial y_r}). \\
\end{align*}

\mypara{Partial derivatives of $\mathbf{\tilde{q}(0)}$}
\begin{align*}
    \frac{\partial \tilde{q}(0)}{\partial q_l} &= \begin{bmatrix}
        \frac{\partial \tilde{\rho}(0)}{\partial \rho_l} & \frac{\partial \tilde{\rho}(0)}{\partial y_l} \\
        \frac{\partial \tilde{y}(0)}{\partial \rho_l} & \frac{\partial \tilde{y}(0)}{\partial y_l} \\
    \end{bmatrix}
\end{align*}
where
\begin{align*}
    \frac{\partial \tilde{\rho}(0)}{\partial \rho_l} &= \frac{1}{\gamma}(\frac{u_l + u_{max}\rho_l^{\gamma}}{(\gamma + 1)u_{max}})^{\frac{1 - \gamma}{\gamma}}(\frac{1}{(\gamma + 1)u_{max}})(\frac{\partial u_l}{\partial \rho_l} + u_{max}\gamma\rho_l^{\gamma - 1}), \\
    \frac{\partial \tilde{\rho}(0)}{\partial y_l} &= \frac{1}{\gamma}(\frac{u_l + u_{max}\rho_l^{\gamma}}{(\gamma + 1)u_{max}})^{\frac{1 - \gamma}{\gamma}}(\frac{1}{(\gamma + 1)u_{max}})(\frac{\partial u_l}{\partial y_l}), \\
    \frac{\partial \tilde{y}(0)}{\partial \rho_l} &= \frac{\partial \tilde{\rho}(0)}{\partial \rho_l}(\tilde{u}(0) - u_{eq}(\tilde{\rho}(0))) + \tilde{\rho}(0)(\frac{\partial \tilde{u}(0)}{\partial \rho_l} - u'_{eq}(\tilde{\rho}(0))\frac{\partial \tilde{\rho}(0)}{\partial \rho_l}), \\
    \frac{\partial \tilde{y}(0)}{\partial y_l} &= \frac{\partial \tilde{\rho}(0)}{\partial y_l}(\tilde{u}(0) - u_{eq}(\tilde{\rho}(0))) + \tilde{\rho}(0)(\frac{\partial \tilde{u}(0)}{\partial y_l} - u'_{eq}(\tilde{\rho}(0))\frac{\partial \tilde{\rho}(0)}{\partial y_l}). \\
\end{align*}
\begin{align*}
    \frac{\partial \tilde{q}(0)}{\partial q_r} &= \begin{bmatrix}
        0 & 0 \\
        0 & 0
    \end{bmatrix}
\end{align*}

%% file: tex/appendix_idm.tex
\subsection{Hyperparameters}
\label{appendix-idm-hyperparams}

In the IDM, following hyperparameters are used to describe different behavioral traits of discrete vehicles~\cite{treiber2000congested}.
\begin{itemize}
    \item $s_{min}$: Minimum desired distance to the leading vehicle.
    \item $T_{pref}$: Desired time to move forward with current speed.
    \item $a_{max}$: Upper bound of the computed acceleration.
    \item $a_{pref}$: Comfortable braking deceleration, positive value.
    \item $v_{targ}$: Target velocity it wants to maintain.
    \item $length_i$: Length of the $i$-th vehicle.
\end{itemize}

\subsection{Differentiation}
\label{appendix-idm-differentiation}

First we compute the gradients of $\mathbf{a_i}(t)$ with respect to $\mathbf{p_i}(t)$, $\mathbf{v_i}(t)$, $\mathbf{p_{h(i)}}(t)$, and $\mathbf{v_{h(i)}}(t)$, where ${h(i)}$-th vehicle is the leading vehicle of $i$-th vehicle.
\begin{align*}
    \frac{\partial \mathbf{a_i}(t)}{\partial \mathbf{p_i}(t)} &= -2a_{max}\frac{{s_{opt}}^{2}}{(p_{h(i)}(t) - p_i(t) - length_{h(i)})^3}, \\
    \frac{\partial \mathbf{a_i}(t)}{\partial \mathbf{p_{h(i)}}(t)} &= 2a_{max}\frac{{s_{opt}}^{2}}{(p_{h(i)}(t) - p_i(t) - length_{h(i)})^3}, \\
    \frac{\partial \mathbf{a_i}(t)}{\partial \mathbf{v_i}(t)} &= a_{max}(-\delta\frac{v_i(t)^{\delta - 1}}{{v_{targ}}^{\delta}} - 2\frac{s_{opt}}{(p_{h(i)}(t) - p_i(t) - length_{h(i)})^2}\frac{\partial s_{opt}}{\partial v_i(t)}), \\
    \frac{\partial \mathbf{a_i}(t)}{\partial \mathbf{v_{h(i)}}(t)} &= a_{max}(-2\frac{s_{opt}}{(p_{h(i)}(t) - p_i(t) - length_{h(i)})^2}\frac{\partial s_{opt}}{\partial v_{h(i)}(t)}), \\
\end{align*}
where
\begin{align*}
    \frac{\partial s_{opt}}{\partial v_i(t)} &= T_{pref} + \frac{2v_i(t) - v_{h(i)}(t)}{2\sqrt{a_{max}a_{pref}}}, \\
    \frac{\partial s_{opt}}{\partial v_{h(i)}(t)} &= \frac{-v_i(t)}{2\sqrt{a_{max}a_{pref}}}.
\end{align*}

Now for arbitrary $i$ and $j$, we can compute the gradient of the state of the $i$-th vehicle with respect to that of the $j$-th vehicle by plugging in the above values. Note that the gradient is non-zero only when $j$ equals to $i$ or $h(i)$.

\begin{equation}
    \frac{\partial \mathbf{p_i}(t+1)}{\partial \mathbf{p_j}(t)} = \begin{cases}
        1 & \mbox{if } j = i \\
        0 & \mbox{if } j = h(i) \\
        0 & \mbox{if } else \\
    \end{cases}
\end{equation}

\begin{equation}
    \frac{\partial \mathbf{p_i}(t+1)}{\partial \mathbf{v_j}(t)} = \begin{cases}
        \Delta{t} & \mbox{if } j = i \\
        0 & \mbox{if } j = h(i) \\
        0 & \mbox{if } else \\
    \end{cases}
\end{equation}

\begin{equation}
    \frac{\partial \mathbf{v_i}(t+1)}{\partial \mathbf{p_j}(t)} = \begin{cases}
        \Delta{t}(-2a_{max}\frac{{s_{opt}}^{2}}{(p_{h(i)}(t) - p_i(t) - length_{h(i)})^3}) & \mbox{if } j = i \\
        \Delta{t}(2a_{max}\frac{{s_{opt}}^{2}}{(p_{h(i)}(t) - p_i(t) - length_{h(i)})^3}) & \mbox{if } j = h(i) \\
        0 & \mbox{if } else \\
    \end{cases}
\end{equation}

\begin{equation}
    \frac{\partial \mathbf{v_i}(t+1)}{\partial \mathbf{v_j}(t)} = \begin{cases}
        1 + \Delta{t}a_{max}(-\delta\frac{{v_i(t)}^{\delta - 1}}{{v_{targ}}^{\delta}} - 2\frac{s_{opt}}{(p_{h(i)}(t) - p_i(t) - length_{h(i)})^2}(T_{pref} + \frac{2v_i(t) - v_j(t)}{2\sqrt{a_{max}a_{pref}}})) & \mbox{if } j = i \\
        \Delta{t}a_{max}(-2\frac{s_{opt}}{(p_{h(i)}(t) - p_i(t) - length_{h(i)})^2}\frac{-v_i(t)}{2\sqrt{a_{max}a_{pref}}}) & \mbox{if } j = h(i) \\
        0 & \mbox{if } else \\
    \end{cases}
\end{equation}

%% file: main.bbl

\begin{thebibliography}{68}


\ifx \showCODEN    \undefined \def \showCODEN     #1{\unskip}     \fi
\ifx \showDOI      \undefined \def \showDOI       #1{#1}\fi
\ifx \showISBNx    \undefined \def \showISBNx     #1{\unskip}     \fi
\ifx \showISBNxiii \undefined \def \showISBNxiii  #1{\unskip}     \fi
\ifx \showISSN     \undefined \def \showISSN      #1{\unskip}     \fi
\ifx \showLCCN     \undefined \def \showLCCN      #1{\unskip}     \fi
\ifx \shownote     \undefined \def \shownote      #1{#1}          \fi
\ifx \showarticletitle \undefined \def \showarticletitle #1{#1}   \fi
\ifx \showURL      \undefined \def \showURL       {\relax}        \fi
\providecommand\bibfield[2]{#2}
\providecommand\bibinfo[2]{#2}
\providecommand\natexlab[1]{#1}
\providecommand\showeprint[2][]{arXiv:#2}

\bibitem[\protect\citeauthoryear{Akhauri, Zheng, and Lin}{Akhauri
  et~al\mbox{.}}{2020}]%
        {akhauri2020enhanced}
\bibfield{author}{\bibinfo{person}{Shivam Akhauri}, \bibinfo{person}{Laura~Y
  Zheng}, {and} \bibinfo{person}{Ming~C Lin}.} \bibinfo{year}{2020}\natexlab{}.
\newblock \showarticletitle{Enhanced transfer learning for autonomous driving
  with systematic accident simulation}. In \bibinfo{booktitle}{\emph{2020
  IEEE/RSJ International Conference on Intelligent Robots and Systems (IROS)}}.
  IEEE, \bibinfo{pages}{5986--5993}.
\newblock


\bibitem[\protect\citeauthoryear{Andelfinger}{Andelfinger}{2021}]%
        {andelfinger2021differentiable}
\bibfield{author}{\bibinfo{person}{Philipp Andelfinger}.}
  \bibinfo{year}{2021}\natexlab{}.
\newblock \showarticletitle{Differentiable Agent-Based Simulation for
  Gradient-Guided Simulation-Based Optimization}. In
  \bibinfo{booktitle}{\emph{Proceedings of the 2021 ACM SIGSIM Conference on
  Principles of Advanced Discrete Simulation}}. \bibinfo{pages}{27--38}.
\newblock


\bibitem[\protect\citeauthoryear{Aw and Rascle}{Aw and Rascle}{2000}]%
        {aw2000resurrection}
\bibfield{author}{\bibinfo{person}{AATM Aw} {and} \bibinfo{person}{Michel
  Rascle}.} \bibinfo{year}{2000}\natexlab{}.
\newblock \showarticletitle{Resurrection of" second order" models of traffic
  flow}.
\newblock \bibinfo{journal}{\emph{SIAM journal on applied mathematics}}
  \bibinfo{volume}{60}, \bibinfo{number}{3} (\bibinfo{year}{2000}),
  \bibinfo{pages}{916--938}.
\newblock


\bibitem[\protect\citeauthoryear{Bando, Hasebe, Nakayama, Shibata, and
  Sugiyama}{Bando et~al\mbox{.}}{1995}]%
        {bando1995dynamical}
\bibfield{author}{\bibinfo{person}{Masako Bando}, \bibinfo{person}{Katsuya
  Hasebe}, \bibinfo{person}{Akihiro Nakayama}, \bibinfo{person}{Akihiro
  Shibata}, {and} \bibinfo{person}{Yuki Sugiyama}.}
  \bibinfo{year}{1995}\natexlab{}.
\newblock \showarticletitle{Dynamical model of traffic congestion and numerical
  simulation}.
\newblock \bibinfo{journal}{\emph{Physical review E}} \bibinfo{volume}{51},
  \bibinfo{number}{2} (\bibinfo{year}{1995}), \bibinfo{pages}{1035}.
\newblock


\bibitem[\protect\citeauthoryear{Bourrel and Lesort}{Bourrel and
  Lesort}{2003}]%
        {bourrel2003mixing}
\bibfield{author}{\bibinfo{person}{Emmanuel Bourrel} {and}
  \bibinfo{person}{Jean-Baptiste Lesort}.} \bibinfo{year}{2003}\natexlab{}.
\newblock \showarticletitle{Mixing microscopic and macroscopic representations
  of traffic flow: Hybrid model based on Lighthill--Whitham--Richards theory}.
\newblock \bibinfo{journal}{\emph{Transportation Research Record}}
  \bibinfo{volume}{1852}, \bibinfo{number}{1} (\bibinfo{year}{2003}),
  \bibinfo{pages}{193--200}.
\newblock


\bibitem[\protect\citeauthoryear{Branicky, Borkar, and Mitter}{Branicky
  et~al\mbox{.}}{1998}]%
        {branicky1998unified}
\bibfield{author}{\bibinfo{person}{Michael~S Branicky},
  \bibinfo{person}{Vivek~S Borkar}, {and} \bibinfo{person}{Sanjoy~K Mitter}.}
  \bibinfo{year}{1998}\natexlab{}.
\newblock \showarticletitle{A unified framework for hybrid control: Model and
  optimal control theory}.
\newblock \bibinfo{journal}{\emph{IEEE transactions on automatic control}}
  \bibinfo{volume}{43}, \bibinfo{number}{1} (\bibinfo{year}{1998}),
  \bibinfo{pages}{31--45}.
\newblock


\bibitem[\protect\citeauthoryear{Cascaval, Shalah, Quinn, Bodik, Agrawala, and
  Schulz}{Cascaval et~al\mbox{.}}{2021}]%
        {cascaval2021differentiable}
\bibfield{author}{\bibinfo{person}{Dan Cascaval}, \bibinfo{person}{Mira
  Shalah}, \bibinfo{person}{Phillip Quinn}, \bibinfo{person}{Rastislav Bodik},
  \bibinfo{person}{Maneesh Agrawala}, {and} \bibinfo{person}{Adriana Schulz}.}
  \bibinfo{year}{2021}\natexlab{}.
\newblock \showarticletitle{Differentiable 3D CAD Programs for Bidirectional
  Editing}.
\newblock \bibinfo{journal}{\emph{arXiv preprint arXiv:2110.01182}}
  (\bibinfo{year}{2021}).
\newblock


\bibitem[\protect\citeauthoryear{Cassidy and Windover}{Cassidy and
  Windover}{1995}]%
        {cassidy1995viethodology}
\bibfield{author}{\bibinfo{person}{Michael~J Cassidy} {and}
  \bibinfo{person}{John~R Windover}.} \bibinfo{year}{1995}\natexlab{}.
\newblock \showarticletitle{v{\'I}ethodology for Assessing Dynamics of Freeway
  Traffic Flow}.
\newblock  (\bibinfo{year}{1995}).
\newblock


\bibitem[\protect\citeauthoryear{Colas, van Toll, Zibrek, Hoyet, Olivier, and
  Pettr{\'e}}{Colas et~al\mbox{.}}{2022}]%
        {colas2022interaction}
\bibfield{author}{\bibinfo{person}{Ad{\`e}le Colas}, \bibinfo{person}{Wouter
  van Toll}, \bibinfo{person}{Katja Zibrek}, \bibinfo{person}{Ludovic Hoyet},
  \bibinfo{person}{Anne-H{\'e}l{\`e}ne Olivier}, {and} \bibinfo{person}{Julien
  Pettr{\'e}}.} \bibinfo{year}{2022}\natexlab{}.
\newblock \showarticletitle{Interaction Fields: Intuitive Sketch-based Steering
  Behaviors for Crowd Simulation}. In \bibinfo{booktitle}{\emph{Computer
  Graphics Forum}}.
\newblock


\bibitem[\protect\citeauthoryear{Daganzo}{Daganzo}{1995}]%
        {daganzo1995requiem}
\bibfield{author}{\bibinfo{person}{Carlos~F Daganzo}.}
  \bibinfo{year}{1995}\natexlab{}.
\newblock \showarticletitle{Requiem for second-order fluid approximations of
  traffic flow}.
\newblock \bibinfo{journal}{\emph{Transportation Research Part B:
  Methodological}} \bibinfo{volume}{29}, \bibinfo{number}{4}
  (\bibinfo{year}{1995}), \bibinfo{pages}{277--286}.
\newblock


\bibitem[\protect\citeauthoryear{de~Avila Belbute-Peres, Smith, Allen,
  Tenenbaum, and Kolter}{de~Avila Belbute-Peres et~al\mbox{.}}{2018}]%
        {de2018end}
\bibfield{author}{\bibinfo{person}{Filipe de Avila Belbute-Peres},
  \bibinfo{person}{Kevin Smith}, \bibinfo{person}{Kelsey Allen},
  \bibinfo{person}{Josh Tenenbaum}, {and} \bibinfo{person}{J~Zico Kolter}.}
  \bibinfo{year}{2018}\natexlab{}.
\newblock \showarticletitle{End-to-end differentiable physics for learning and
  control}.
\newblock \bibinfo{journal}{\emph{Advances in neural information processing
  systems}}  \bibinfo{volume}{31} (\bibinfo{year}{2018}),
  \bibinfo{pages}{7178--7189}.
\newblock


\bibitem[\protect\citeauthoryear{Dosovitskiy, Ros, Codevilla, Lopez, and
  Koltun}{Dosovitskiy et~al\mbox{.}}{2017}]%
        {Dosovitskiy17}
\bibfield{author}{\bibinfo{person}{Alexey Dosovitskiy}, \bibinfo{person}{German
  Ros}, \bibinfo{person}{Felipe Codevilla}, \bibinfo{person}{Antonio Lopez},
  {and} \bibinfo{person}{Vladlen Koltun}.} \bibinfo{year}{2017}\natexlab{}.
\newblock \showarticletitle{{CARLA}: {An} Open Urban Driving Simulator}. In
  \bibinfo{booktitle}{\emph{Proceedings of the 1st Annual Conference on Robot
  Learning}}. \bibinfo{pages}{1--16}.
\newblock


\bibitem[\protect\citeauthoryear{Du, Wu, Ma, Wah, Spielberg, Rus, and
  Matusik}{Du et~al\mbox{.}}{2021}]%
        {du2021diffpd}
\bibfield{author}{\bibinfo{person}{Tao Du}, \bibinfo{person}{Kui Wu},
  \bibinfo{person}{Pingchuan Ma}, \bibinfo{person}{Sebastien Wah},
  \bibinfo{person}{Andrew Spielberg}, \bibinfo{person}{Daniela Rus}, {and}
  \bibinfo{person}{Wojciech Matusik}.} \bibinfo{year}{2021}\natexlab{}.
\newblock \showarticletitle{{DiffPD}: Differentiable Projective Dynamics with
  Contact}.
\newblock \bibinfo{journal}{\emph{arXiv:2101.05917}} (\bibinfo{year}{2021}).
\newblock


\bibitem[\protect\citeauthoryear{Du, Wu, Spielberg, Matusik, Zhu, and
  Sifakis}{Du et~al\mbox{.}}{2020}]%
        {du2020stokes}
\bibfield{author}{\bibinfo{person}{Tao Du}, \bibinfo{person}{Kui Wu},
  \bibinfo{person}{Andrew Spielberg}, \bibinfo{person}{Wojciech Matusik},
  \bibinfo{person}{Bo Zhu}, {and} \bibinfo{person}{Eftychios Sifakis}.}
  \bibinfo{year}{2020}\natexlab{}.
\newblock \showarticletitle{Functional Optimization of Fluidic Devices with
  Differentiable Stokes Flow}.
\newblock \bibinfo{journal}{\emph{{ACM} Trans. Graph.}} (\bibinfo{year}{2020}).
\newblock


\bibitem[\protect\citeauthoryear{Eom and Kim}{Eom and Kim}{2020}]%
        {eom2020traffic}
\bibfield{author}{\bibinfo{person}{Myungeun Eom} {and}
  \bibinfo{person}{Byung-In Kim}.} \bibinfo{year}{2020}\natexlab{}.
\newblock \showarticletitle{The traffic signal control problem for
  intersections: a review}.
\newblock \bibinfo{journal}{\emph{European transport research review}}
  \bibinfo{volume}{12}, \bibinfo{number}{1} (\bibinfo{year}{2020}),
  \bibinfo{pages}{1--20}.
\newblock


\bibitem[\protect\citeauthoryear{Fierro, Das, Kumar, and Ostrowski}{Fierro
  et~al\mbox{.}}{2001}]%
        {fierro2001hybrid}
\bibfield{author}{\bibinfo{person}{Rafael Fierro}, \bibinfo{person}{Aveek~K
  Das}, \bibinfo{person}{Vijay Kumar}, {and} \bibinfo{person}{James~P
  Ostrowski}.} \bibinfo{year}{2001}\natexlab{}.
\newblock \showarticletitle{Hybrid control of formations of robots}. In
  \bibinfo{booktitle}{\emph{Proceedings 2001 ICRA. IEEE International
  Conference on Robotics and Automation (Cat. No. 01CH37164)}},
  Vol.~\bibinfo{volume}{1}. IEEE, \bibinfo{pages}{157--162}.
\newblock


\bibitem[\protect\citeauthoryear{Gazis, Herman, and Potts}{Gazis
  et~al\mbox{.}}{1959}]%
        {gazis1959car}
\bibfield{author}{\bibinfo{person}{Denos~C Gazis}, \bibinfo{person}{Robert
  Herman}, {and} \bibinfo{person}{Renfrey~B Potts}.}
  \bibinfo{year}{1959}\natexlab{}.
\newblock \showarticletitle{Car-following theory of steady-state traffic flow}.
\newblock \bibinfo{journal}{\emph{Operations research}} \bibinfo{volume}{7},
  \bibinfo{number}{4} (\bibinfo{year}{1959}), \bibinfo{pages}{499--505}.
\newblock


\bibitem[\protect\citeauthoryear{Gazis, Herman, and Rothery}{Gazis
  et~al\mbox{.}}{1961}]%
        {gazis1961nonlinear}
\bibfield{author}{\bibinfo{person}{Denos~C Gazis}, \bibinfo{person}{Robert
  Herman}, {and} \bibinfo{person}{Richard~W Rothery}.}
  \bibinfo{year}{1961}\natexlab{}.
\newblock \showarticletitle{Nonlinear follow-the-leader models of traffic
  flow}.
\newblock \bibinfo{journal}{\emph{Operations research}} \bibinfo{volume}{9},
  \bibinfo{number}{4} (\bibinfo{year}{1961}), \bibinfo{pages}{545--567}.
\newblock


\bibitem[\protect\citeauthoryear{Geilinger, Hahn, Zehnder, B{\"{a}}cher,
  Thomaszewski, and Coros}{Geilinger et~al\mbox{.}}{2020}]%
        {Geilinger2020add}
\bibfield{author}{\bibinfo{person}{Moritz Geilinger}, \bibinfo{person}{David
  Hahn}, \bibinfo{person}{Jonas Zehnder}, \bibinfo{person}{Moritz
  B{\"{a}}cher}, \bibinfo{person}{Bernhard Thomaszewski}, {and}
  \bibinfo{person}{Stelian Coros}.} \bibinfo{year}{2020}\natexlab{}.
\newblock \showarticletitle{{ADD}: Analytically differentiable dynamics for
  multi-body systems with frictional contact}.
\newblock \bibinfo{journal}{\emph{ACM Transactions on Graphics (TOG)}}
  \bibinfo{volume}{39}, \bibinfo{number}{6} (\bibinfo{year}{2020}).
\newblock


\bibitem[\protect\citeauthoryear{Gipps}{Gipps}{1981}]%
        {gipps1981behavioural}
\bibfield{author}{\bibinfo{person}{Peter~G Gipps}.}
  \bibinfo{year}{1981}\natexlab{}.
\newblock \showarticletitle{A behavioural car-following model for computer
  simulation}.
\newblock \bibinfo{journal}{\emph{Transportation Research Part B:
  Methodological}} \bibinfo{volume}{15}, \bibinfo{number}{2}
  (\bibinfo{year}{1981}), \bibinfo{pages}{105--111}.
\newblock


\bibitem[\protect\citeauthoryear{Golas, Narain, Sewall, Krajcevski, Dubey, and
  Lin}{Golas et~al\mbox{.}}{2012}]%
        {golas2012large}
\bibfield{author}{\bibinfo{person}{Abhinav Golas}, \bibinfo{person}{Rahul
  Narain}, \bibinfo{person}{Jason Sewall}, \bibinfo{person}{Pavel Krajcevski},
  \bibinfo{person}{Pradeep Dubey}, {and} \bibinfo{person}{Ming Lin}.}
  \bibinfo{year}{2012}\natexlab{}.
\newblock \showarticletitle{Large-scale fluid simulation using
  velocity-vorticity domain decomposition}.
\newblock \bibinfo{journal}{\emph{ACM Transactions on Graphics (TOG)}}
  \bibinfo{volume}{31}, \bibinfo{number}{6} (\bibinfo{year}{2012}),
  \bibinfo{pages}{1--9}.
\newblock


\bibitem[\protect\citeauthoryear{Haarnoja, Zhou, Abbeel, and Levine}{Haarnoja
  et~al\mbox{.}}{2018}]%
        {haarnoja2018soft}
\bibfield{author}{\bibinfo{person}{Tuomas Haarnoja}, \bibinfo{person}{Aurick
  Zhou}, \bibinfo{person}{Pieter Abbeel}, {and} \bibinfo{person}{Sergey
  Levine}.} \bibinfo{year}{2018}\natexlab{}.
\newblock \showarticletitle{Soft actor-critic: Off-policy maximum entropy deep
  reinforcement learning with a stochastic actor}. In
  \bibinfo{booktitle}{\emph{International conference on machine learning}}.
  PMLR, \bibinfo{pages}{1861--1870}.
\newblock


\bibitem[\protect\citeauthoryear{H{\"a}drich, Banuti, Pa{\l}ubicki, Pirk, and
  Michels}{H{\"a}drich et~al\mbox{.}}{2021}]%
        {hadrich2021fire}
\bibfield{author}{\bibinfo{person}{Torsten H{\"a}drich},
  \bibinfo{person}{Daniel~T Banuti}, \bibinfo{person}{Wojtek Pa{\l}ubicki},
  \bibinfo{person}{S{\"o}ren Pirk}, {and} \bibinfo{person}{Dominik~L Michels}.}
  \bibinfo{year}{2021}\natexlab{}.
\newblock \showarticletitle{Fire in paradise: mesoscale simulation of
  wildfires}.
\newblock \bibinfo{journal}{\emph{ACM Transactions on Graphics (TOG)}}
  \bibinfo{volume}{40}, \bibinfo{number}{4} (\bibinfo{year}{2021}),
  \bibinfo{pages}{1--15}.
\newblock


\bibitem[\protect\citeauthoryear{Hansen}{Hansen}{2006}]%
        {hansen2006cma}
\bibfield{author}{\bibinfo{person}{Nikolaus Hansen}.}
  \bibinfo{year}{2006}\natexlab{}.
\newblock \showarticletitle{The CMA evolution strategy: a comparing review}.
\newblock \bibinfo{journal}{\emph{Towards a new evolutionary computation}}
  (\bibinfo{year}{2006}), \bibinfo{pages}{75--102}.
\newblock


\bibitem[\protect\citeauthoryear{He, Xiang, Zhao, and Wang}{He
  et~al\mbox{.}}{2020}]%
        {he2020informative}
\bibfield{author}{\bibinfo{person}{Feixiang He}, \bibinfo{person}{Yuanhang
  Xiang}, \bibinfo{person}{Xi Zhao}, {and} \bibinfo{person}{He Wang}.}
  \bibinfo{year}{2020}\natexlab{}.
\newblock \showarticletitle{Informative scene decomposition for crowd analysis,
  comparison and simulation guidance}.
\newblock \bibinfo{journal}{\emph{ACM Transactions on Graphics (TOG)}}
  \bibinfo{volume}{39}, \bibinfo{number}{4} (\bibinfo{year}{2020}),
  \bibinfo{pages}{50--1}.
\newblock


\bibitem[\protect\citeauthoryear{Heiden, Macklin, Narang, Fox, Garg, and
  Ramos}{Heiden et~al\mbox{.}}{2021}]%
        {heiden2021disect}
\bibfield{author}{\bibinfo{person}{Eric Heiden}, \bibinfo{person}{Miles
  Macklin}, \bibinfo{person}{Yashraj~S Narang}, \bibinfo{person}{Dieter Fox},
  \bibinfo{person}{Animesh Garg}, {and} \bibinfo{person}{Fabio Ramos}.}
  \bibinfo{year}{2021}\natexlab{}.
\newblock \showarticletitle{{DiSECt: A Differentiable Simulation Engine for
  Autonomous Robotic Cutting}}. In \bibinfo{booktitle}{\emph{Proceedings of
  Robotics: Science and Systems}}. \bibinfo{address}{Virtual}.
\newblock
\urldef\tempurl%
\url{https://doi.org/10.15607/RSS.2021.XVII.067}
\showDOI{\tempurl}


\bibitem[\protect\citeauthoryear{Holl, Koltun, Um, and Thuerey}{Holl
  et~al\mbox{.}}{2020}]%
        {holl2020phiflow}
\bibfield{author}{\bibinfo{person}{Philipp Holl}, \bibinfo{person}{Vladlen
  Koltun}, \bibinfo{person}{Kiwon Um}, {and} \bibinfo{person}{Nils Thuerey}.}
  \bibinfo{year}{2020}\natexlab{}.
\newblock \showarticletitle{phiflow: A differentiable pde solving framework for
  deep learning via physical simulations}. In \bibinfo{booktitle}{\emph{NeurIPS
  Workshop}}.
\newblock


\bibitem[\protect\citeauthoryear{Hu, Anderson, Li, Sun, Carr, Ragan{-}Kelley,
  and Durand}{Hu et~al\mbox{.}}{2020}]%
        {Hu2019:ICLR}
\bibfield{author}{\bibinfo{person}{Yuanming Hu}, \bibinfo{person}{Luke
  Anderson}, \bibinfo{person}{Tzu-Mao Li}, \bibinfo{person}{Qi Sun},
  \bibinfo{person}{Nathan Carr}, \bibinfo{person}{Jonathan Ragan{-}Kelley},
  {and} \bibinfo{person}{Fr{\'{e}}do Durand}.} \bibinfo{year}{2020}\natexlab{}.
\newblock \showarticletitle{{DiffTaichi}: Differentiable Programming for
  Physical Simulation}. In \bibinfo{booktitle}{\emph{ICLR}}.
\newblock


\bibitem[\protect\citeauthoryear{Ishiwaka, Zeng, Eastman, Kakazu, Gross,
  Mizutani, and Nakada}{Ishiwaka et~al\mbox{.}}{2021}]%
        {ishiwaka2021foids}
\bibfield{author}{\bibinfo{person}{Yuko Ishiwaka}, \bibinfo{person}{Xiao~S
  Zeng}, \bibinfo{person}{Michael~Lee Eastman}, \bibinfo{person}{Sho Kakazu},
  \bibinfo{person}{Sarah Gross}, \bibinfo{person}{Ryosuke Mizutani}, {and}
  \bibinfo{person}{Masaki Nakada}.} \bibinfo{year}{2021}\natexlab{}.
\newblock \showarticletitle{Foids: bio-inspired fish simulation for generating
  synthetic datasets}.
\newblock \bibinfo{journal}{\emph{ACM Transactions on Graphics (TOG)}}
  \bibinfo{volume}{40}, \bibinfo{number}{6} (\bibinfo{year}{2021}),
  \bibinfo{pages}{1--15}.
\newblock


\bibitem[\protect\citeauthoryear{Jiang, Wu, and Zhu}{Jiang
  et~al\mbox{.}}{2001}]%
        {jiang2001full}
\bibfield{author}{\bibinfo{person}{Rui Jiang}, \bibinfo{person}{Qingsong Wu},
  {and} \bibinfo{person}{Zuojin Zhu}.} \bibinfo{year}{2001}\natexlab{}.
\newblock \showarticletitle{Full velocity difference model for a car-following
  theory}.
\newblock \bibinfo{journal}{\emph{Physical Review E}} \bibinfo{volume}{64},
  \bibinfo{number}{1} (\bibinfo{year}{2001}), \bibinfo{pages}{017101}.
\newblock


\bibitem[\protect\citeauthoryear{Kesting, Treiber, and Helbing}{Kesting
  et~al\mbox{.}}{2007}]%
        {kesting2007general}
\bibfield{author}{\bibinfo{person}{Arne Kesting}, \bibinfo{person}{Martin
  Treiber}, {and} \bibinfo{person}{Dirk Helbing}.}
  \bibinfo{year}{2007}\natexlab{}.
\newblock \showarticletitle{General lane-changing model MOBIL for car-following
  models}.
\newblock \bibinfo{journal}{\emph{Transportation Research Record}}
  \bibinfo{volume}{1999}, \bibinfo{number}{1} (\bibinfo{year}{2007}),
  \bibinfo{pages}{86--94}.
\newblock


\bibitem[\protect\citeauthoryear{Kraft et~al\mbox{.}}{Kraft
  et~al\mbox{.}}{1988}]%
        {kraft1988software}
\bibfield{author}{\bibinfo{person}{Dieter Kraft} {et~al\mbox{.}}}
  \bibinfo{year}{1988}\natexlab{}.
\newblock \showarticletitle{A software package for sequential quadratic
  programming}.
\newblock  (\bibinfo{year}{1988}).
\newblock


\bibitem[\protect\citeauthoryear{LeVeque et~al\mbox{.}}{LeVeque
  et~al\mbox{.}}{2002}]%
        {leveque2002finite}
\bibfield{author}{\bibinfo{person}{Randall~J LeVeque} {et~al\mbox{.}}}
  \bibinfo{year}{2002}\natexlab{}.
\newblock \bibinfo{booktitle}{\emph{Finite volume methods for hyperbolic
  problems}}. Vol.~\bibinfo{volume}{31}.
\newblock \bibinfo{publisher}{Cambridge university press}.
\newblock


\bibitem[\protect\citeauthoryear{Li, Aittala, Durand, and Lehtinen}{Li
  et~al\mbox{.}}{2018}]%
        {li2018diff}
\bibfield{author}{\bibinfo{person}{Tzu-Mao Li}, \bibinfo{person}{Miika
  Aittala}, \bibinfo{person}{Fr{\'{e}}do Durand}, {and} \bibinfo{person}{Jaakko
  Lehtinen}.} \bibinfo{year}{2018}\natexlab{}.
\newblock \showarticletitle{Differentiable Monte Carlo ray tracing through edge
  sampling}.
\newblock \bibinfo{journal}{\emph{{ACM} Trans. Graph.}} \bibinfo{volume}{37},
  \bibinfo{number}{6} (\bibinfo{year}{2018}).
\newblock


\bibitem[\protect\citeauthoryear{Li, Du, Wu, Xu, and Matusik}{Li
  et~al\mbox{.}}{2022}]%
        {li2022diffcloth}
\bibfield{author}{\bibinfo{person}{Yifei Li}, \bibinfo{person}{Tao Du},
  \bibinfo{person}{Kui Wu}, \bibinfo{person}{Jie Xu}, {and}
  \bibinfo{person}{Wojciech Matusik}.} \bibinfo{year}{2022}\natexlab{}.
\newblock \showarticletitle{DiffCloth: Differentiable Cloth Simulation with Dry
  Frictional Contact}.
\newblock \bibinfo{journal}{\emph{ACM Trans. Graph.}} (\bibinfo{year}{2022}).
\newblock


\bibitem[\protect\citeauthoryear{Lieberman and Rathi}{Lieberman and
  Rathi}{1997}]%
        {lieberman1997traffic}
\bibfield{author}{\bibinfo{person}{Edward Lieberman} {and}
  \bibinfo{person}{Ajay~K Rathi}.} \bibinfo{year}{1997}\natexlab{}.
\newblock \showarticletitle{Traffic simulation}.
\newblock \bibinfo{journal}{\emph{Traffic flow theory}} (\bibinfo{year}{1997}).
\newblock


\bibitem[\protect\citeauthoryear{Lighthill and Whitham}{Lighthill and
  Whitham}{1955}]%
        {lighthill1955kinematic}
\bibfield{author}{\bibinfo{person}{Michael~James Lighthill} {and}
  \bibinfo{person}{Gerald~Beresford Whitham}.} \bibinfo{year}{1955}\natexlab{}.
\newblock \showarticletitle{On kinematic waves II. A theory of traffic flow on
  long crowded roads}.
\newblock \bibinfo{journal}{\emph{Proceedings of the Royal Society of London.
  Series A. Mathematical and Physical Sciences}} \bibinfo{volume}{229},
  \bibinfo{number}{1178} (\bibinfo{year}{1955}), \bibinfo{pages}{317--345}.
\newblock


\bibitem[\protect\citeauthoryear{Lillicrap, Hunt, Pritzel, Heess, Erez, Tassa,
  Silver, and Wierstra}{Lillicrap et~al\mbox{.}}{2015}]%
        {lillicrap2015continuous}
\bibfield{author}{\bibinfo{person}{Timothy~P Lillicrap},
  \bibinfo{person}{Jonathan~J Hunt}, \bibinfo{person}{Alexander Pritzel},
  \bibinfo{person}{Nicolas Heess}, \bibinfo{person}{Tom Erez},
  \bibinfo{person}{Yuval Tassa}, \bibinfo{person}{David Silver}, {and}
  \bibinfo{person}{Daan Wierstra}.} \bibinfo{year}{2015}\natexlab{}.
\newblock \showarticletitle{Continuous control with deep reinforcement
  learning}.
\newblock \bibinfo{journal}{\emph{arXiv preprint arXiv:1509.02971}}
  (\bibinfo{year}{2015}).
\newblock


\bibitem[\protect\citeauthoryear{Lopez, Behrisch, Bieker-Walz, Erdmann,
  Fl{\"o}tter{\"o}d, Hilbrich, L{\"u}cken, Rummel, Wagner, and
  Wie{\ss}ner}{Lopez et~al\mbox{.}}{2018}]%
        {lopez2018microscopic}
\bibfield{author}{\bibinfo{person}{Pablo~Alvarez Lopez},
  \bibinfo{person}{Michael Behrisch}, \bibinfo{person}{Laura Bieker-Walz},
  \bibinfo{person}{Jakob Erdmann}, \bibinfo{person}{Yun-Pang
  Fl{\"o}tter{\"o}d}, \bibinfo{person}{Robert Hilbrich},
  \bibinfo{person}{Leonhard L{\"u}cken}, \bibinfo{person}{Johannes Rummel},
  \bibinfo{person}{Peter Wagner}, {and} \bibinfo{person}{Evamarie
  Wie{\ss}ner}.} \bibinfo{year}{2018}\natexlab{}.
\newblock \showarticletitle{Microscopic traffic simulation using sumo}. In
  \bibinfo{booktitle}{\emph{2018 21st International Conference on Intelligent
  Transportation Systems (ITSC)}}. IEEE, \bibinfo{pages}{2575--2582}.
\newblock


\bibitem[\protect\citeauthoryear{Ma, Du, Zhang, Wu, Spielberg, Katzschmann, and
  Matusik}{Ma et~al\mbox{.}}{2021}]%
        {ma2021diffaqua}
\bibfield{author}{\bibinfo{person}{Pingchuan Ma}, \bibinfo{person}{Tao Du},
  \bibinfo{person}{John~Z Zhang}, \bibinfo{person}{Kui Wu},
  \bibinfo{person}{Andrew Spielberg}, \bibinfo{person}{Robert~K Katzschmann},
  {and} \bibinfo{person}{Wojciech Matusik}.} \bibinfo{year}{2021}\natexlab{}.
\newblock \showarticletitle{Diffaqua: A differentiable computational design
  pipeline for soft underwater swimmers with shape interpolation}.
\newblock \bibinfo{journal}{\emph{ACM Transactions on Graphics (TOG)}}
  \bibinfo{volume}{40}, \bibinfo{number}{4} (\bibinfo{year}{2021}),
  \bibinfo{pages}{1--14}.
\newblock


\bibitem[\protect\citeauthoryear{Magne, Rabut, and Gabard}{Magne
  et~al\mbox{.}}{2000}]%
        {magne2000towards}
\bibfield{author}{\bibinfo{person}{Laurent Magne}, \bibinfo{person}{Sylvestre
  Rabut}, {and} \bibinfo{person}{Jean-Fran{\c{c}}ois Gabard}.}
  \bibinfo{year}{2000}\natexlab{}.
\newblock \showarticletitle{Towards an hybrid macro-micro traffic flow
  simulation model}. In \bibinfo{booktitle}{\emph{INFORMS spring 2000
  meeting}}.
\newblock


\bibitem[\protect\citeauthoryear{Mammar, Mammar, and Lebacque}{Mammar
  et~al\mbox{.}}{2006}]%
        {mammar2006highway}
\bibfield{author}{\bibinfo{person}{Salim Mammar}, \bibinfo{person}{Sa{\"\i}d
  Mammar}, {and} \bibinfo{person}{Jean-Patrick Lebacque}.}
  \bibinfo{year}{2006}\natexlab{}.
\newblock \showarticletitle{Highway traffic hybrid macro-micro simulation
  model}.
\newblock \bibinfo{journal}{\emph{IFAC Proceedings Volumes}}
  \bibinfo{volume}{39}, \bibinfo{number}{12} (\bibinfo{year}{2006}),
  \bibinfo{pages}{627--632}.
\newblock


\bibitem[\protect\citeauthoryear{Mohamed and Mohamad}{Mohamed and
  Mohamad}{2010}]%
        {mohamed2010review}
\bibfield{author}{\bibinfo{person}{Khaled~M Mohamed} {and} \bibinfo{person}{AA
  Mohamad}.} \bibinfo{year}{2010}\natexlab{}.
\newblock \showarticletitle{A review of the development of hybrid
  atomistic--continuum methods for dense fluids}.
\newblock \bibinfo{journal}{\emph{Microfluidics and Nanofluidics}}
  \bibinfo{volume}{8}, \bibinfo{number}{3} (\bibinfo{year}{2010}),
  \bibinfo{pages}{283--302}.
\newblock


\bibitem[\protect\citeauthoryear{Mora, Peychev, Ha, Vechev, and Coros}{Mora
  et~al\mbox{.}}{2021}]%
        {mora2021pods}
\bibfield{author}{\bibinfo{person}{Miguel Angel~Zamora Mora},
  \bibinfo{person}{Momchil~P Peychev}, \bibinfo{person}{Sehoon Ha},
  \bibinfo{person}{Martin Vechev}, {and} \bibinfo{person}{Stelian Coros}.}
  \bibinfo{year}{2021}\natexlab{}.
\newblock \showarticletitle{PODS: Policy Optimization via Differentiable
  Simulation}. In \bibinfo{booktitle}{\emph{International Conference on Machine
  Learning}}. PMLR, \bibinfo{pages}{7805--7817}.
\newblock


\bibitem[\protect\citeauthoryear{Narain, Golas, Curtis, and Lin}{Narain
  et~al\mbox{.}}{2009}]%
        {narain2009aggregate}
\bibfield{author}{\bibinfo{person}{Rahul Narain}, \bibinfo{person}{Abhinav
  Golas}, \bibinfo{person}{Sean Curtis}, {and} \bibinfo{person}{Ming~C Lin}.}
  \bibinfo{year}{2009}\natexlab{}.
\newblock \showarticletitle{Aggregate dynamics for dense crowd simulation}.
\newblock In \bibinfo{booktitle}{\emph{ACM SIGGRAPH Asia 2009 papers}}.
  \bibinfo{pages}{1--8}.
\newblock


\bibitem[\protect\citeauthoryear{Narain, Golas, and Lin}{Narain
  et~al\mbox{.}}{2010}]%
        {narain2010free}
\bibfield{author}{\bibinfo{person}{Rahul Narain}, \bibinfo{person}{Abhinav
  Golas}, {and} \bibinfo{person}{Ming~C Lin}.} \bibinfo{year}{2010}\natexlab{}.
\newblock \showarticletitle{Free-flowing granular materials with two-way solid
  coupling}.
\newblock In \bibinfo{booktitle}{\emph{ACM SIGGRAPH Asia 2010 papers}}.
  \bibinfo{pages}{1--10}.
\newblock


\bibitem[\protect\citeauthoryear{Nelder and Mead}{Nelder and Mead}{1965}]%
        {nelder1965simplex}
\bibfield{author}{\bibinfo{person}{John~A Nelder} {and} \bibinfo{person}{Roger
  Mead}.} \bibinfo{year}{1965}\natexlab{}.
\newblock \showarticletitle{A simplex method for function minimization}.
\newblock \bibinfo{journal}{\emph{The computer journal}} \bibinfo{volume}{7},
  \bibinfo{number}{4} (\bibinfo{year}{1965}), \bibinfo{pages}{308--313}.
\newblock


\bibitem[\protect\citeauthoryear{Newell}{Newell}{1961}]%
        {newell1961nonlinear}
\bibfield{author}{\bibinfo{person}{Gordon~Frank Newell}.}
  \bibinfo{year}{1961}\natexlab{}.
\newblock \showarticletitle{Nonlinear effects in the dynamics of car
  following}.
\newblock \bibinfo{journal}{\emph{Operations research}} \bibinfo{volume}{9},
  \bibinfo{number}{2} (\bibinfo{year}{1961}), \bibinfo{pages}{209--229}.
\newblock


\bibitem[\protect\citeauthoryear{Nimier{-}David, Vicini, Zeltner, and
  Jakob}{Nimier{-}David et~al\mbox{.}}{2019}]%
        {david2019mitsuba}
\bibfield{author}{\bibinfo{person}{Merlin Nimier{-}David},
  \bibinfo{person}{Delio Vicini}, \bibinfo{person}{Tizian Zeltner}, {and}
  \bibinfo{person}{Wenzel Jakob}.} \bibinfo{year}{2019}\natexlab{}.
\newblock \showarticletitle{Mitsuba 2: A retargetable forward and inverse
  renderer}.
\newblock \bibinfo{journal}{\emph{ACM Transactions on Graphics (TOG)}}
  \bibinfo{volume}{38}, \bibinfo{number}{6} (\bibinfo{year}{2019}).
\newblock


\bibitem[\protect\citeauthoryear{Paszke, Gross, Massa, Lerer, Bradbury, Chanan,
  Killeen, Lin, Gimelshein, Antiga, et~al\mbox{.}}{Paszke
  et~al\mbox{.}}{2019}]%
        {paszke2019pytorch}
\bibfield{author}{\bibinfo{person}{Adam Paszke}, \bibinfo{person}{Sam Gross},
  \bibinfo{person}{Francisco Massa}, \bibinfo{person}{Adam Lerer},
  \bibinfo{person}{James Bradbury}, \bibinfo{person}{Gregory Chanan},
  \bibinfo{person}{Trevor Killeen}, \bibinfo{person}{Zeming Lin},
  \bibinfo{person}{Natalia Gimelshein}, \bibinfo{person}{Luca Antiga},
  {et~al\mbox{.}}} \bibinfo{year}{2019}\natexlab{}.
\newblock \showarticletitle{Pytorch: An imperative style, high-performance deep
  learning library}.
\newblock \bibinfo{journal}{\emph{Advances in neural information processing
  systems}}  \bibinfo{volume}{32} (\bibinfo{year}{2019}),
  \bibinfo{pages}{8026--8037}.
\newblock


\bibitem[\protect\citeauthoryear{Payne}{Payne}{1971}]%
        {payne1971model}
\bibfield{author}{\bibinfo{person}{Harold~J Payne}.}
  \bibinfo{year}{1971}\natexlab{}.
\newblock \showarticletitle{Model of freeway traffic and control}.
\newblock \bibinfo{journal}{\emph{Mathematical Model of Public System}}
  (\bibinfo{year}{1971}), \bibinfo{pages}{51--61}.
\newblock


\bibitem[\protect\citeauthoryear{Qiao, Liang, Koltun, and Lin}{Qiao
  et~al\mbox{.}}{2020}]%
        {qiao2020scalable}
\bibfield{author}{\bibinfo{person}{Yi-Ling Qiao}, \bibinfo{person}{Junbang
  Liang}, \bibinfo{person}{Vladlen Koltun}, {and} \bibinfo{person}{Ming~C.
  Lin}.} \bibinfo{year}{2020}\natexlab{}.
\newblock \showarticletitle{Scalable Differentiable Physics for Learning and
  Control}. In \bibinfo{booktitle}{\emph{ICML}}.
\newblock


\bibitem[\protect\citeauthoryear{Qiao, Liang, Koltun, and Lin}{Qiao
  et~al\mbox{.}}{2021a}]%
        {Qiao2021Differentiable}
\bibfield{author}{\bibinfo{person}{Yi-Ling Qiao}, \bibinfo{person}{Junbang
  Liang}, \bibinfo{person}{Vladlen Koltun}, {and} \bibinfo{person}{Ming~C.
  Lin}.} \bibinfo{year}{2021}\natexlab{a}.
\newblock \showarticletitle{Differentiable Simulation of Soft Multi-body
  Systems}. In \bibinfo{booktitle}{\emph{Conference on Neural Information
  Processing Systems (NeurIPS)}}.
\newblock


\bibitem[\protect\citeauthoryear{Qiao, Liang, Koltun, and Lin}{Qiao
  et~al\mbox{.}}{2021b}]%
        {qiao2021Efficient}
\bibfield{author}{\bibinfo{person}{Yi-Ling Qiao}, \bibinfo{person}{Junbang
  Liang}, \bibinfo{person}{Vladlen Koltun}, {and} \bibinfo{person}{Ming~C.
  Lin}.} \bibinfo{year}{2021}\natexlab{b}.
\newblock \showarticletitle{Efficient Differentiable Simulation of Articulated
  Bodies}. In \bibinfo{booktitle}{\emph{ICML}}.
\newblock


\bibitem[\protect\citeauthoryear{Richards}{Richards}{1956}]%
        {richards1956shock}
\bibfield{author}{\bibinfo{person}{Paul~I Richards}.}
  \bibinfo{year}{1956}\natexlab{}.
\newblock \showarticletitle{Shock waves on the highway}.
\newblock \bibinfo{journal}{\emph{Operations research}} \bibinfo{volume}{4},
  \bibinfo{number}{1} (\bibinfo{year}{1956}), \bibinfo{pages}{42--51}.
\newblock


\bibitem[\protect\citeauthoryear{Schulman, Wolski, Dhariwal, Radford, and
  Klimov}{Schulman et~al\mbox{.}}{2017}]%
        {schulman2017proximal}
\bibfield{author}{\bibinfo{person}{John Schulman}, \bibinfo{person}{Filip
  Wolski}, \bibinfo{person}{Prafulla Dhariwal}, \bibinfo{person}{Alec Radford},
  {and} \bibinfo{person}{Oleg Klimov}.} \bibinfo{year}{2017}\natexlab{}.
\newblock \showarticletitle{Proximal policy optimization algorithms}.
\newblock \bibinfo{journal}{\emph{arXiv preprint arXiv:1707.06347}}
  (\bibinfo{year}{2017}).
\newblock


\bibitem[\protect\citeauthoryear{Sewall, Wilkie, and Lin}{Sewall
  et~al\mbox{.}}{2011}]%
        {sewall2011interactive}
\bibfield{author}{\bibinfo{person}{Jason Sewall}, \bibinfo{person}{David
  Wilkie}, {and} \bibinfo{person}{Ming~C Lin}.}
  \bibinfo{year}{2011}\natexlab{}.
\newblock \showarticletitle{Interactive hybrid simulation of large-scale
  traffic}. In \bibinfo{booktitle}{\emph{Proceedings of the 2011 SIGGRAPH Asia
  Conference}}. \bibinfo{pages}{1--12}.
\newblock


\bibitem[\protect\citeauthoryear{Sewall, Wilkie, Merrell, and Lin}{Sewall
  et~al\mbox{.}}{2010}]%
        {sewall2010continuum}
\bibfield{author}{\bibinfo{person}{Jason Sewall}, \bibinfo{person}{David
  Wilkie}, \bibinfo{person}{Paul Merrell}, {and} \bibinfo{person}{Ming~C Lin}.}
  \bibinfo{year}{2010}\natexlab{}.
\newblock \showarticletitle{Continuum traffic simulation}. In
  \bibinfo{booktitle}{\emph{Computer Graphics Forum}},
  Vol.~\bibinfo{volume}{29}. Wiley Online Library, \bibinfo{pages}{439--448}.
\newblock


\bibitem[\protect\citeauthoryear{Sewall}{Sewall}{2011}]%
        {sewall2011efficient}
\bibfield{author}{\bibinfo{person}{Jason~Douglas Sewall}.}
  \bibinfo{year}{2011}\natexlab{}.
\newblock \emph{\bibinfo{title}{Efficient, scalable traffic and compressible
  fluid simulations using hyperbolic models}}.
\newblock \bibinfo{thesistype}{Ph.D. Dissertation}. \bibinfo{school}{The
  University of North Carolina at Chapel Hill}.
\newblock


\bibitem[\protect\citeauthoryear{Shen, Yin, Shao, Wang, Jiang, Lan, and
  Zhou}{Shen et~al\mbox{.}}{2021}]%
        {shen2021high}
\bibfield{author}{\bibinfo{person}{Siyuan Shen}, \bibinfo{person}{Yang Yin},
  \bibinfo{person}{Tianjia Shao}, \bibinfo{person}{He Wang},
  \bibinfo{person}{Chenfanfu Jiang}, \bibinfo{person}{Lei Lan}, {and}
  \bibinfo{person}{Kun Zhou}.} \bibinfo{year}{2021}\natexlab{}.
\newblock \showarticletitle{High-order differentiable autoencoder for nonlinear
  model reduction}.
\newblock \bibinfo{journal}{\emph{arXiv preprint arXiv:2102.11026}}
  (\bibinfo{year}{2021}).
\newblock


\bibitem[\protect\citeauthoryear{Takahashi, Liang, Qiao, and Lin}{Takahashi
  et~al\mbox{.}}{2021}]%
        {takahashi2021differentiable}
\bibfield{author}{\bibinfo{person}{Tetsuya Takahashi}, \bibinfo{person}{Junbang
  Liang}, \bibinfo{person}{Yi-Ling Qiao}, {and} \bibinfo{person}{Ming~C Lin}.}
  \bibinfo{year}{2021}\natexlab{}.
\newblock \showarticletitle{Differentiable Fluids with Solid Coupling for
  Learning and Control}. In \bibinfo{booktitle}{\emph{Proceedings of the AAAI
  Conference on Artificial Intelligence}}, Vol.~\bibinfo{volume}{35}.
  \bibinfo{pages}{6138--6146}.
\newblock


\bibitem[\protect\citeauthoryear{Treiber, Hennecke, and Helbing}{Treiber
  et~al\mbox{.}}{2000}]%
        {treiber2000congested}
\bibfield{author}{\bibinfo{person}{Martin Treiber}, \bibinfo{person}{Ansgar
  Hennecke}, {and} \bibinfo{person}{Dirk Helbing}.}
  \bibinfo{year}{2000}\natexlab{}.
\newblock \showarticletitle{Congested traffic states in empirical observations
  and microscopic simulations}.
\newblock \bibinfo{journal}{\emph{Physical review E}} \bibinfo{volume}{62},
  \bibinfo{number}{2} (\bibinfo{year}{2000}), \bibinfo{pages}{1805}.
\newblock


\bibitem[\protect\citeauthoryear{Treuille, Cooper, and Popovi{\'c}}{Treuille
  et~al\mbox{.}}{2006}]%
        {treuille2006continuum}
\bibfield{author}{\bibinfo{person}{Adrien Treuille}, \bibinfo{person}{Seth
  Cooper}, {and} \bibinfo{person}{Zoran Popovi{\'c}}.}
  \bibinfo{year}{2006}\natexlab{}.
\newblock \showarticletitle{Continuum crowds}.
\newblock \bibinfo{journal}{\emph{ACM Transactions on Graphics (TOG)}}
  \bibinfo{volume}{25}, \bibinfo{number}{3} (\bibinfo{year}{2006}),
  \bibinfo{pages}{1160--1168}.
\newblock


\bibitem[\protect\citeauthoryear{Wei, Zheng, Yao, and Li}{Wei
  et~al\mbox{.}}{2018}]%
        {wei2018intellilight}
\bibfield{author}{\bibinfo{person}{Hua Wei}, \bibinfo{person}{Guanjie Zheng},
  \bibinfo{person}{Huaxiu Yao}, {and} \bibinfo{person}{Zhenhui Li}.}
  \bibinfo{year}{2018}\natexlab{}.
\newblock \showarticletitle{Intellilight: A reinforcement learning approach for
  intelligent traffic light control}. In \bibinfo{booktitle}{\emph{Proceedings
  of the 24th ACM SIGKDD International Conference on Knowledge Discovery \&
  Data Mining}}. \bibinfo{pages}{2496--2505}.
\newblock


\bibitem[\protect\citeauthoryear{Whitham}{Whitham}{2011}]%
        {whitham2011linear}
\bibfield{author}{\bibinfo{person}{Gerald~Beresford Whitham}.}
  \bibinfo{year}{2011}\natexlab{}.
\newblock \bibinfo{booktitle}{\emph{Linear and nonlinear waves}}.
  Vol.~\bibinfo{volume}{42}.
\newblock \bibinfo{publisher}{John Wiley \& Sons}.
\newblock


\bibitem[\protect\citeauthoryear{Yu and Krstic}{Yu and Krstic}{2019}]%
        {yu2019traffic}
\bibfield{author}{\bibinfo{person}{Huan Yu} {and} \bibinfo{person}{Miroslav
  Krstic}.} \bibinfo{year}{2019}\natexlab{}.
\newblock \showarticletitle{Traffic congestion control for Aw--Rascle--Zhang
  model}.
\newblock \bibinfo{journal}{\emph{Automatica}}  \bibinfo{volume}{100}
  (\bibinfo{year}{2019}), \bibinfo{pages}{38--51}.
\newblock


\bibitem[\protect\citeauthoryear{Zhang}{Zhang}{2002}]%
        {zhang2002non}
\bibfield{author}{\bibinfo{person}{H~Michael Zhang}.}
  \bibinfo{year}{2002}\natexlab{}.
\newblock \showarticletitle{A non-equilibrium traffic model devoid of gas-like
  behavior}.
\newblock \bibinfo{journal}{\emph{Transportation Research Part B:
  Methodological}} \bibinfo{volume}{36}, \bibinfo{number}{3}
  (\bibinfo{year}{2002}), \bibinfo{pages}{275--290}.
\newblock


\bibitem[\protect\citeauthoryear{Zheng, Ma, and Wang}{Zheng
  et~al\mbox{.}}{2017}]%
        {zheng2017consensus}
\bibfield{author}{\bibinfo{person}{Yuanshi Zheng}, \bibinfo{person}{Jingying
  Ma}, {and} \bibinfo{person}{Long Wang}.} \bibinfo{year}{2017}\natexlab{}.
\newblock \showarticletitle{Consensus of hybrid multi-agent systems}.
\newblock \bibinfo{journal}{\emph{IEEE transactions on neural networks and
  learning systems}} \bibinfo{volume}{29}, \bibinfo{number}{4}
  (\bibinfo{year}{2017}), \bibinfo{pages}{1359--1365}.
\newblock


\end{thebibliography}
